\documentclass[times]{weauth} \usepackage{subfig} \usepackage{moreverb}

\newcommand*\patchAmsMathEnvironmentForLineno[1]{%
  \expandafter\let\csname old#1\expandafter\endcsname\csname #1\endcsname
  \expandafter\let\csname oldend#1\expandafter\endcsname\csname end#1\endcsname
  \renewenvironment{#1}%
     {\linenomath\csname old#1\endcsname}%
     {\csname oldend#1\endcsname\endlinenomath}}%
\newcommand*\patchBothAmsMathEnvironmentsForLineno[1]{%
  \patchAmsMathEnvironmentForLineno{#1}%
  \patchAmsMathEnvironmentForLineno{#1*}}%
\AtBeginDocument{%
\patchBothAmsMathEnvironmentsForLineno{equation}%
\patchBothAmsMathEnvironmentsForLineno{align}%
\patchBothAmsMathEnvironmentsForLineno{flalign}%
\patchBothAmsMathEnvironmentsForLineno{alignat}%
\patchBothAmsMathEnvironmentsForLineno{gather}%
\patchBothAmsMathEnvironmentsForLineno{multline}%
}

\newcommand\BibTeX{{\rmfamily B\kern-.05em \textsc{i\kern-.025em b}\kern-.08em T\kern-.1667em\lower.7ex\hbox{E}\kern-.125emX}}

\def\volumeyear{2017}

\usepackage{lineno}

\usepackage{graphicx}
\usepackage{epstopdf}

\usepackage{hyperref}
\hypersetup{
    bookmarks=true,         % show bookmarks bar?
    unicode=false,          % non-Latin characters in Acrobat?s bookmarks
    pdftoolbar=true,        % show Acrobat?s toolbar?
    pdfmenubar=true,        % show Acrobat?s menu?
    pdffitwindow=false,     % window fit to page when opened
    pdftitle={},    % title
    pdfauthor={},     % author
    pdfsubject={},   % subject of the document
    pdfcreator={},   % creator of the document
    pdfproducer={},  % producer of the document
    pdfkeywords={}, % list of keywords
    pdfnewwindow=true,      % links in new window
    colorlinks=false,       % false: boxed links; true: colored links
    linkcolor=red,          % color of internal links
    citecolor=green,        % color of links to bibliography
    filecolor=magenta,      % color of file links
    urlcolor=cyan           % color of external links
}

\usepackage{nomencl}
\makenomenclature

\begin{document}

\runningheads{Xiaolei Yang and Fotis Sotiropoulos}{A new class of actuator surface models for wind turbines}

\articletype{RESEARCH ARTICLE}

\title{A new class of actuator surface models for wind turbines}

\author{Xiaolei Yang$^1$, Fotis Sotiropoulos$^{1}$}

\address{$^1$ Department of Civil Engineering, College of Engineering and Applied Sciences, Stony Brook University, Stony Brook, NY 11794, USA}

%\author{}
%\address{}

\corraddr{Fotis Sotiropoulos, Department of Civil Engineering, College of Engineering and Applied Sciences, Stony Brook University, Stony Brook, NY 11794, USA \\ Email address: fotis.sotiropoulos@stonybrook.edu.}

\begin{abstract} 
Actuator line model has been widely employed in wind turbine simulations. However, the standard actuator line model does not include a model for the turbine nacelle which can significantly impact turbine wake characteristics as shown in the literature (e.g. Kang, Yang and Sotiropoulos, Journal of Fluid Mechanics 744 (2014): 376-403; Viola et al., Journal of Fluid Mechanics 750 (2014): R1; Foti et al., Physical Review Fluids 1 (2016), 044407). Another disadvantage of the standard actuator line model is that more geometrical features of turbine blades cannot be resolved on a finer mesh. To alleviate these disadvantages of the standard model,  we develop a new class of actuator surface models for turbine blades and nacelle to take into account more geometrical details of turbine blades and include the effect of turbine nacelle.  
In the actuator surface model for blade, the aerodynamic forces calculated using the blade element method are distributed from the surface formed by the foil chords at different radial locations. In the actuator surface model for nacelle, the forces are distributed from the actual nacelle surface with the normal force component computed in the same way as in the direct forcing immersed boundary method and the tangential force component computed using a friction coefficient and a reference velocity of the incoming flow. The actuator surface model for nacelle is evaluated by simulating the flow over periodically placed nacelles. Both the actuator surface simulation and the wall-resolved large-eddy simulation are carried out.  The comparison shows that the actuator surface model is able to give acceptable results especially at far wake locations on a very coarse mesh. \textcolor[rgb]{0,0,0}{It is noted that although this model is employed for the turbine nacelle in this work, it is also applicable to other bluff bodies. }
The capability of the actuator surface model in predicting turbine wakes is assessed by simulating the flow over the MEXICO (Model experiments in Controlled Conditions) turbine and the hydrokinetic turbine of Kang, Yang and Sotiropoulos (Journal of Fluid Mechanics 744 (2014): 376-403). Comparisons of the computed results with measurements show that the proposed actuator surface model is able to predict the tip vortices, turbulence statistics and meandering of turbine wake with good accuracy. 
\end{abstract}

\keywords{Turbine parameterization; Nacelle model; Wake meandering}

\maketitle
\renewcommand{\nomname}{List of symbols}
\printnomenclature
%\linenumbers 

\section{Introduction}
\label{sec:introduction}
Using the current state-of-the-art supercomputers, it is still impossible to simulate wind farms resolving all the relevant length scales from the thickness of the boundary layer over a wind turbine blade ($\approx$0.016 m at the trailing edge estimated using the equation ${\delta}/{c}=0.37\left( {U_{\infty}c}/{\nu}\right)^{-0.2}$\cite{schlichting2003boundary} with $\nu=1.5\times10^{-5}$ $m^2/s$, and airfoil chord $c=1$ m and $U_{\infty}=100$ $m/s$ at the blade tip for a MW scale wind turbine with rotor diameter approximately 100 $m$, tip speed ratio 10 and incoming wind speed 10 $m/s$) to the thickness of an atmospheric boundary layer ($\approx$1000 $m$~\cite{tennekes1972first}). Different rotor models including the actuator disk model~\cite{calaf2010large, yang2012computational, yang2014large}, the actuator line model~\cite{sorensen2002numerical, yang2015large} and the actuator surface model~\cite{shen2009actuator} have been developed in the literature to parametrize the interaction between the turbine blades and the incoming flow without directly resolving the boundary layer flow over turbine blades. 

In the actuator disk model~\cite{froude1889part} the wind turbine rotor is modelled by a circular permeable disk. The impacts of incoming wind on wind turbines are taken into account by uniformly distributed forces on the disk. In the conventional actuator disk model, the forces are only applied along the turbine axial direction, which can be calculated either from one-dimensional momentum theory with a prescribed induction factor~\cite{calaf2010large} or from the blade-element momentum theory~\cite{masson2001aerodynamic}. In the work by Wu and Port{\'e}-Agel~\cite{wu2011large}, the rotation was taken into account in the actuator disk model. It was found that the predictions from the actuator disk model with rotation show better agreement with the wind tunnel measurements reported in~\cite{chamorro2010effects}. 

The actuator line model provides a more accurate approach to incorporate the rotational effect, finite blade number effect and the effect of non-uniform force distribution in the azimuthal direction. In this model, the turbine blade is represented by a rotating line with distributed forces, which are calculated from a blade element approach~\cite{hansen2015aerodynamics} combined with tabulated drag and lift coefficients of the foils. The employed drag and lift coefficients are usually from two-dimensional numerical simulations or experiments without considering the rotational effects. In order to take into account the three-dimensional rotational effects, corrections on the two-dimensional force coefficients, such as those proposed in~\cite{snel1993sectional}, are usually employed. As discussed by Shen et al.~\cite{shen2005tip}, the lift needs to approach zero as the blade tip is approached. However, the lift on the tip predicted by the blade element approach is usually not zero especially for a pitched blade or a chambered foil. A tip-loss correction in addition to the correction for the three-dimensional effect is needed in order to have a correct lift distribution in the region near the blade tip as suggested by Shen et al.~\cite{shen2005tip2}. 
%
%Prediction capabilities of  the actuator type models have been evaluated at both field-scale~\cite{ammara2002viscous, jimenez2007advances, troldborg2010numerical, troldborg2011numerical} and laboratory-scale~\cite{wu2011large, porte2011large, shen2012actuator,  wu2013simulation, krogstad2013blind, pierella2014blind, yang2015large, krogstad2015blind, nilsson2015validation}. 
%Although overall good agreements with experimental results have been observed for the actuator line model,
There are two major limitations of the standard actuator line model: 
%\begin{enumerate}
{i) The lack of an effective nacelle model.} Nacelle induced coherent structures were shown to have a significant impact on turbine wake characteristics, such as velocity deficit in the near wake~\cite{kang2014onset, yang2015large, tossas2016wind} and wake meandering at far wake locations for model wind turbines~\cite{viola2014prediction, howard2015statistics, foti2016wake} and hydrokinetic turbines~\cite{kang2014onset}. Both geometry-resolving, wall-modelled large-eddy simulations and actuator type large-eddy simulations were carried out in~\cite{kang2014onset}. It is shown that the actuator line model without a nacelle model cannot accurately capture the wake meandering of the hydrokinetic turbine, and underpredicts the turbulence intensity at far wake locations. 
This was attributed to the fact that the presence of the nacelle induces spiral vortex breakdown of the hub vortex causing a large-scale energetic coherent vortex that precesses and expands radially outward to intercept and energize the tip shear layer at three to four rotor diameters downwind of the turbine~\cite{kang2014onset}.
The nacelle can be represented using a permeable disk with a specified drag coefficient.  However, Yang et al.~\cite{yang2015large} showed that such a simple nacelle model is not able to take into account the nacelle effect accurately. The computed results from Tossas et al.~\cite{tossas2016wind} indicated that with their nacelle and tower model the velocity deficit in the near wake is predicted more accurately. The improvement on the prediction of the turbulence intensity, on the other hand, is very limited. 
{ii) A finer mesh cannot resolve more geometrical features of the turbine blade.} The results of Yang et al.~\cite{yang2015large} showed that grid-independent results cannot be obtained when the mesh is refined in the actuator line model simulations. \textcolor[rgb]{0,0,0}{This is because more physics cannot be resolved on a finer grid in the state-of-the-art actuator type models. The accuracy of the computed results mainly depends on the parameterizations employed in actuator type models.}
%\end{enumerate}

The objective of this paper is to develop a new class of actuator surface models for turbine blade and nacelle to take into account nacelle geometry and better incorporating the geometrical effect of turbine blade. In the actuator surface model for blade, the forces are still computed using the blade element method but are distributed on the surface formed by the foil chords at different radial locations. In the actuator surface model for the nacelle, the normal forces are computed by satisfying the non-penetration boundary conditions as in the direct forcing immersed boundary method~\cite{sotiropoulos2014immersed, uhlmann2005immersed, yang2009smoothing}. The tangential forces, on the other hand, are calculated using a friction coefficient and a reference velocity of the incoming flow. 
More details on the proposed actuator surface models are presented in Sec.~\ref{sec:ASM} followed by a brief description of the governing equations and flow solver in Sec.~\ref{sec:flowsolver}. Some cases for validating the proposed actuator surface model are then reported in Sec.~\ref{sec: testcases}. Finally, conclusions are drawn in Sec.~\ref{sec:Conclusions}.
\section{Actuator surface models}
\label{sec:ASM}
In this section, we present the proposed actuator surface models for the blade and nacelle. In both actuator surface models, we have two sets of independent meshes, i.e., the background Cartesian grid for the flow with its coordinate denoted by $\mathbf{x}$ ($x$, $y$, $z$ or $x_1$, $x_2$, $x_3$), and the Lagrangian grid following the actuator surfaces with its coordinate denoted by $\mathbf{X}$ ($X$, $Y$, $Z$ or $X_1$, $X_2$, $X_3$). For both models, the smoothed discrete delta function in~\cite{yang2009smoothing} is employed as the kernel function for transferring information between the two meshes. The major difference between the actuator surface model for blade and the actuator surface model for nacelle is the way how the forces on the actuator surfaces are calculated. In Sec.~\ref{sec:actuator surface model for blades} we present the force calculation and distribution methods of the actuator surface model for blade. And in Sec.~\ref{sec:actuator surface model for nacelle} we introduce the forces calculation method for the actuator surface model for nacelle. 
\subsection{Actuator surface model for blade}
\label{sec:actuator surface model for blades}
\begin{figure}
  \centerline{\includegraphics[width=5.0in]{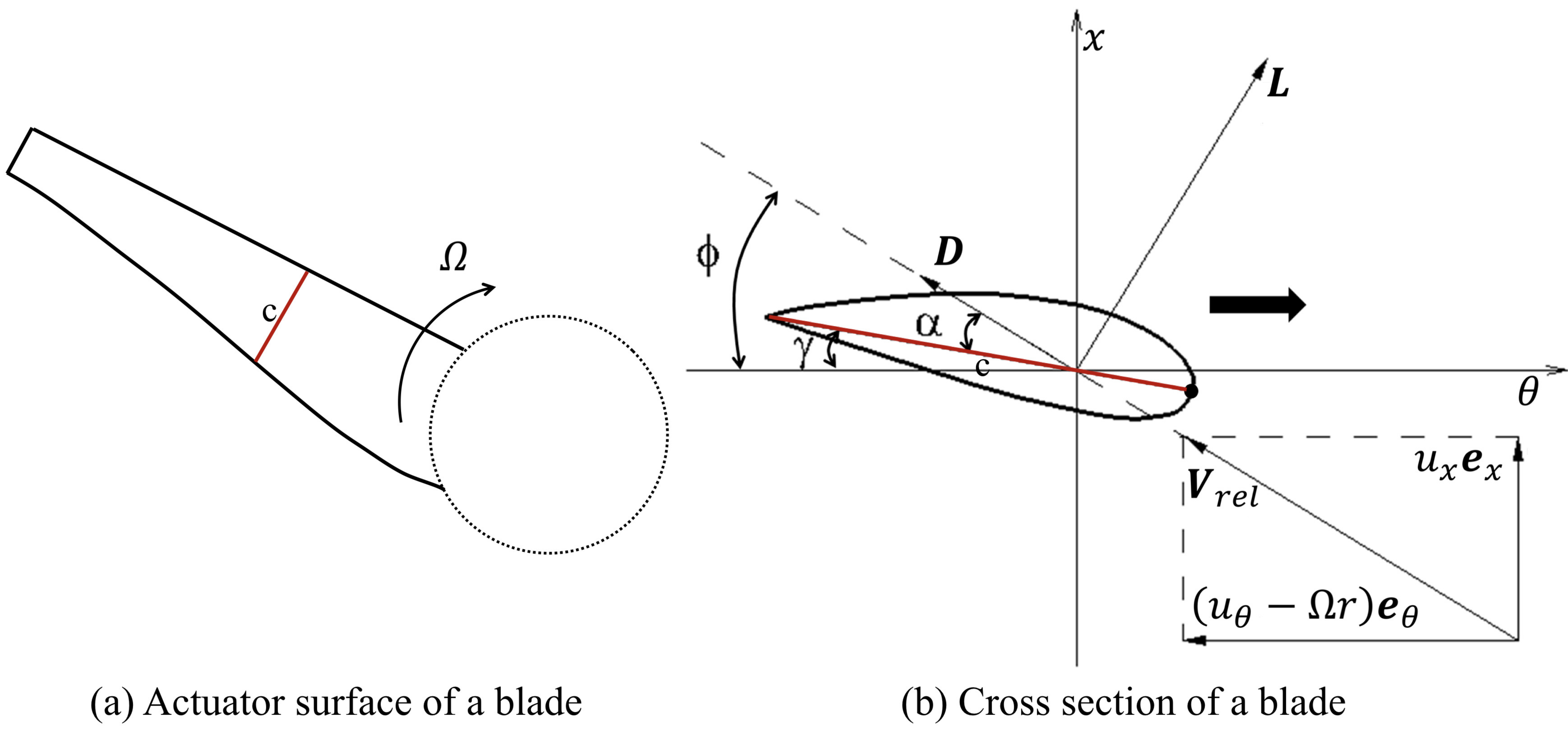}}
  \caption{A schematic of the actuator surface model for blade. The lift and drag forces calculated using the blade element method are distributed over the actuator surface formed by chord lines of a blade. }
\label{fig:AS_schematic}
\end{figure}
In the actuator surface model for blade, the blade geometry is represented by a surface formed by the chord lines at different radial locations of a blade as shown in Fig.~\ref{fig:AS_schematic}(a). The forces are calculated in the same way as in the actuator line model. The lift ($\mathbf{L}$) and drag ($\mathbf{D}$) at each radial location are calculated by
\begin{equation}
\mathbf{L}=\frac{1}{2} \rho C_L c |\mathbf{V}_{rel}|^2\mathbf{e}_{L}
\end{equation}
and
\begin{equation}
\mathbf{D}=\frac{1}{2} \rho C_D c |\mathbf{V}_{rel}|^2\mathbf{e}_{D}, 
\end{equation}
where $C_L$ and $C_D$ are the lift and drag coefficients, $c$ is the chord length, $\mathbf{V}_{rel}$ is the relative incoming velocity, and $\mathbf{e}_{L}$ and $\mathbf{e}_{D}$ are the unit vectors in the directions of lift and drag, respectively. The $C_L$ and $C_D$ are functions of the Reynolds number and the angle of attack. The angle of attack $\alpha$ shown in Fig.~\ref{fig:AS_schematic}(b) is computed at each radial location by 
\begin{equation}
\alpha=\phi-\gamma, 
\end{equation}
where $\phi=-\tan^{-1}(u_x/(u_\theta-\Omega r))$, and $\gamma$ is the angle including the blade twist and the blade pitch, the latter of which is specified to a given value instead of from a turbine control algorithm. 
The relative incoming velocity $\mathbf{V}_{rel}$ at each radial location is computed by 
\begin{equation}\label{eq:Vrel}
\mathbf{V}_{rel} = u_x\mathbf{e}_x+(u_\theta-\Omega r)\mathbf{e}_\theta,
\end{equation}
where $\Omega$ is the rotational speed of the rotor, $\mathbf{e}_x$ and $\mathbf{e}_\theta$ are the unit vectors in the axial and azimuthal directions, respectively. The $u_x$ and $u_\theta$ are the axial and azimuthal components of the flow velocity averaged over the chord for each radial locations, which are computed using
\begin{equation}
u_x=\frac{1}{c}\int_{c}{\mathbf{u}\left(\mathbf{X}\right)\cdot \mathbf{e}_x}ds
\end{equation}
and
\begin{equation}
u_\theta=\frac{1}{c}\int_{c}{\mathbf{u}\left(\mathbf{X}\right)\cdot \mathbf{e}_\theta}ds,
\end{equation}
where $\mathbf{X}$ denote the coordinates of the grid points on the actuator surfaces. 
Generally the grid points on the actuator surfaces do not coincide with any background nodes. We employ a smoothed discrete delta function (i.e. the smoothed four-point cosine function) proposed by Yang et al.~\cite{yang2009smoothing} to interpolate $\mathbf{u}\left(\mathbf{X}\right)$ from the values on the background grid nodes as follows:
\begin{equation}\label{eq:vel-intp}
\mathbf{u}\left(\mathbf{X}\right) = \sum_{\mathbf{x} \in g_{\mathbf{x}}} {\mathbf{u}(\mathbf{x})\delta_h\left(\mathbf{x}-\mathbf{X}\right)}V({\mathbf{x}}),
\end{equation}
where $\mathbf{x}$ are the coordinates of the background grid nodes, $g_{\mathbf{x}}$ is the set of the background grid cells, $V=h_xh_yh_z$ ($h_x$, $h_y$ and $h_z$ are the grid spacings in the $x$, $y$ and $z$ directions, respectively) is the volume of the background grid cell, $\delta_h\left( \mathbf{x}-\mathbf{X} \right)=\frac{1}{V}\phi\left(\frac{x-X}{h_x} \right)\phi\left(\frac{y-Y}{h_y} \right)\phi\left(\frac{z-Z}{h_z} \right)$ is the discrete delta function, and 
$\phi$ is the smoothed four-point cosine function~\cite{yang2009smoothing}, which is expressed as
\begin{equation}\label{eq:phi_4sc}
\phi(r)= \left\{
\begin{array}{llr}
\frac{1}{4}+\frac{\sin\left({\pi}\left(2|r|+1\right)/4\right) }{2\pi}- \frac{\sin\left({\pi}\left(2|r|-1\right)/4\right)}{2 \pi}, && |r| \leq 1.5, \\
\frac{5}{8}-\frac{|r|}{4}-\frac{\sin\left({\pi}\left(2|r|-1\right)/4\right) }{2 \pi}, && 1.5 \leq |r| \leq 2.5,\\
0, && 2.5 \leq |r|, \\
\end{array} 
\right.
\end{equation}
in which $r_i=(x_i-X_i)/h_i $ ($i$=1, 2, 3). 

The blade rotation causes the stall delay phenomenon at the inboard sections of the blade, which increases the lift coefficients and decreases the drag coefficients as compared with the corresponding two-dimensional airfoil data. To account for such three-dimensional rotational effect, the stall delay model developed by Du and Selig~\cite{du19983} is employed to correct the lift and drag coefficients from two-dimensional experiments or numerical simulations. In Du and Selig's model, the corrected lift and drag coefficients ($C_{L,3D}$ and $C_{D,3D}$) are calculated as follows:
\begin{equation}\label{eg:CL_DS}
C_{L,3D}=C_{L,2D}+f_{L}\left(C_{L,p}-C_{L,2D} \right),
\end{equation}
and 
\begin{equation}\label{eq:CD_DS}
C_{D,3D}=C_{D,2D}-f_{D}\left(C_{D,2D}-C_{D,0} \right),
\end{equation}
where $C_{L,p} = 2\pi\left(\alpha - \alpha_0\right)$, $C_{D,0} $ the two-dimensional drag coefficient at zero angle of attack, and the correction functions $f_L$ and $f_D$ are determined by 
\begin{equation}
f_{L}=\frac{1}{2\pi}\left(\frac{1.6(c/r)a-(c/r)^{\frac{d}{\Lambda}\frac{R}{r}}}{0.1267b+(c/r)^{\frac{d}{\Lambda}\frac{R}{r}}}-1 \right),
\end{equation}
and
\begin{equation}
f_{D}=\frac{1}{2\pi}\left(\frac{1.6(c/r)a-(c/r)^{\frac{d}{2\Lambda}\frac{R}{r}}}{0.1267b+(c/r)^{\frac{d}{2\Lambda}\frac{R}{r}}}-1 \right),
\end{equation}
respectively, where $\Lambda=\Omega R/\sqrt{U^2+(\Omega R)^2}$, $U$ is the incoming wind speed, $R$ is the rotor radius, $a$, $b$ and $d$ are the empirical correction factors. In this work, a, b and d are equal to 1 as in Du and Selig's paper~\cite{du19983}.  

Non-zero force can exist at the blade tip when the pitch angle is nonzero or a chambered foil is used~\cite{shen2005tip}. This is in contradiction with the physical understanding that the force should tend to zero at the tip due to pressure equalization as discussed by Shen et al.~\cite{shen2005tip}. To correct this non-physical force behaviour, the tip-loss correction proposed by Shen et al.~\cite{shen2005tip2, shen2005tip} is applied to the drag and lift coefficients computed from Eqs.~(\ref{eg:CL_DS}) and (\ref{eq:CD_DS}). With the tip loss correction, the employed $C_D$ and $C_L$ are calculated as 
\begin{equation}
C_L=F_1C_{L,3D},
\end{equation}
and 
\begin{equation}
C_D=F_1C_{D,3D},
\end{equation}
where 
\begin{equation}
F_1=\frac{2}{\pi}\cos^{-1}\left(\exp\left(-g\frac{B(R-r)}{2r\sin \phi} \right) \right),
\end{equation}
in which $B$ is the number of blades, $g$ is computed by
\begin{equation}
g=\exp\left(-c_1 \left(B\Omega R /U_\infty -c_2 \right) \right)+c_3,
\end{equation}
where $c_1$, $c_2$ and $c_3$ are the correction coefficients, which are equal to 0.125, 21 and 0.1 as in~\cite{shen2005tip2}, respectively. 

After calculating the lift ($\mathbf{L}$) and drag ($\mathbf{D}$), the force per unit area on the actuator surface at each radial location is then calculated as 
\begin{equation}
\mathbf{f}(\mathbf{X})=\left(\mathbf{L}+\mathbf{D}\right)/c.
\end{equation} 
Note that the above expression essentially means that the lift and drag on the blade are uniformly distributed in the chordwise direction. 
To calculate the forces on the background mesh for the flowfield, the computed forces on the actuator surfaces are then distributed to the background grid nodes as follows: 
\begin{equation}\label{eq:force-dist}
\mathbf{f}(\mathbf{x})=-\sum_{\mathbf{X} \in g_{\mathbf{X}}} \mathbf{f}({\mathbf{X}}) \delta_h\left(\mathbf{x}-\mathbf{X}\right)A(\mathbf{X}),
\end{equation} 
where $g_{\mathbf{X}}$ is the set of the actuator surface grid cells and $A$ is the area of the actuator surface grid cell. The same discrete delta function as in Eq.~(\ref{eq:vel-intp}) is employed. It is noted the negative sign is because that $\mathbf{f}(\mathbf{x})$ represents the forces of the actuator surfaces acting on the flow, while $\mathbf{f}({\mathbf{X}}) $ denotes the forces of the flow acting on the actuator surfaces. 
\subsection{Actuator surface model for nacelle}
\label{sec:actuator surface model for nacelle}
Resolving the boundary layer over a nacelle of an utility scale wind turbine, which becomes almost impossible for wind farm scale simulations, requires a much finer mesh as compared with the mesh for resolving the thickness of an atmospheric boundary layer. To take into account the nacelle effects on a relatively coarse mesh, in this section we present an actuator surface model for parametrizing the nacelle effect without directly simulating the boundary layer flows over the nacelle. 
\begin{figure}
  \centerline{\includegraphics[width=\textwidth]{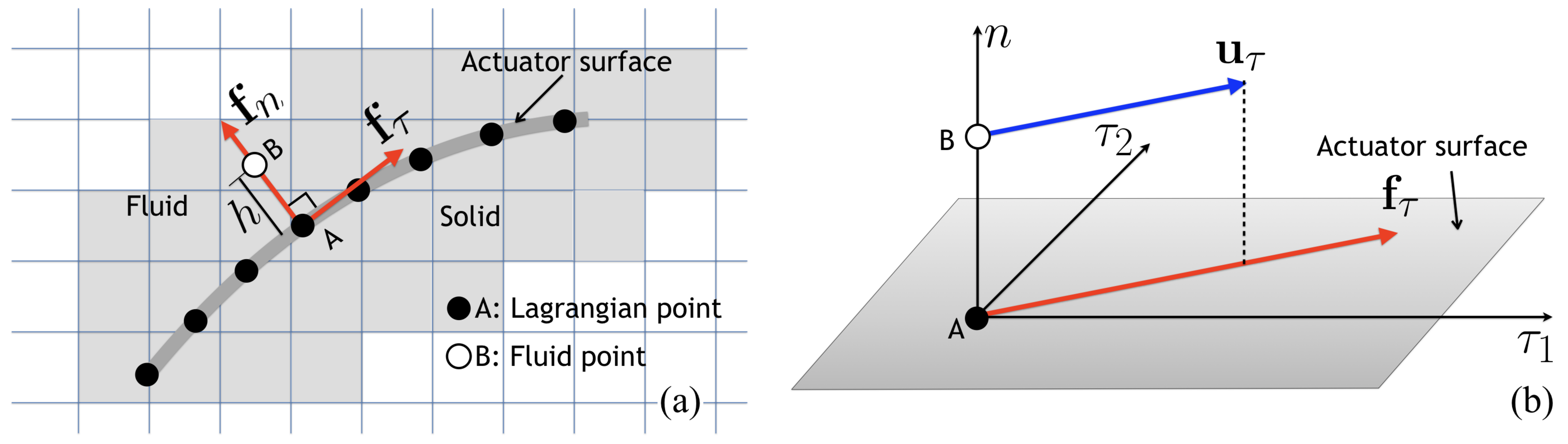}}
  \caption{A schematic of the actuator surface model for nacelle. As shown in (a), the normal force $\mathbf{f}_n$ at the Lagrangian point A is computed by Eq.~(\ref{eq:Fn_ASN}) with the estimated velocity $\mathbf{\tilde{u}}(\mathbf{X})$ at point A interpolated from the surrounding Cartesian grid nodes using Eq.~(\ref{eq:vel-intp}). The magnitude of the tangential force $\mathbf{f}_\tau$ is computed by Eq.~(\ref{eq:Ft_ASN}) using the incoming streamwise velocity $U$ and a friction coefficient given by Eq.~(\ref{eq:cf}). The direction of the tangential force is assumed to be the same as the tangential velocity at point B, which is interpolated from the surrounding Cartesian grid nodes using  Eq.~(\ref{eq:vel-intp}). The computed forces are distributed to the surrounding Cartesian grid nodes (the gray area shown in (a)) using Eq.~(\ref{eq:force-dist}).}
\label{fig:nacellemodel}
\end{figure}

In this model, the nacelle geometry is represented by the actual surface of the nacelle with distributed forces. As shown in Fig.~\ref{fig:nacellemodel}, the force acting on the actuator surface is decomposed into the normal component and the tangential component. From a physical point of view, the actuator surface of the nacelle cannot act as a sink or source of mass not matter how the boundary layer over the nacelle is modelled. Based on this consideration, the normal component of the force on the actuator surface is computed by satisfying the non-penetration constraint. The non-penetration constraint can be met by directly reconstructing the wall-normal velocity near the actuator surface as in the sharp interface immersed boundary method~\cite{sotiropoulos2014immersed, gilmanov2005hybrid, yang2006embedded, ge2007numerical}. However, a physically reasonable interpolation scheme is difficult to implement because of the very coarse gird employed in the actuator type simulations.  Moreover direct velocity reconstruction cannot ensure a smooth representation of the nacelle geometry because of the very coarse grid. In this work we represent the nacelle geometry as a diffused interface the same as in the direct forcing immersed boundary method~\cite{uhlmann2005immersed, yang2009smoothing, wang2011immersed}. In this actuator surface model for nacelle, the normal component of the force acting on the surface per unit area is calculated in the same way as in the direct forcing immersed boundary method, which is expressed as follows:
\begin{equation} \label{eq:Fn_ASN}
\mathbf{f}_n(\mathbf{X})=\frac{h\left(-\mathbf{u}^d(\mathbf{X})+\mathbf{\tilde{u}}(\mathbf{X})\right)\cdot \mathbf{e}_n(\mathbf{X})}{\Delta t}\mathbf{e}_n(\mathbf{X}),
\end{equation}
where $\mathbf{u}^d(\mathbf{X})$ is the desired velocity on the nacelle surface, $\mathbf{e}_n(\mathbf{X})$ is the unit vector in the normal direction of the nacelle surface, $h=(h_x h_y h_z)^{1/3}$ is the length scale of the local background grid spacing (different values of $h$ have been tested without noticing any significant differences), $\mathbf{\tilde{u}}(\mathbf{X})$ is the estimated velocity on the actuator surface, which is interpolated from the corresponding estimated velocity on the background mesh using Eq.~(\ref{eq:vel-intp}) with the $\mathbf{\tilde{u}}(\mathbf{x})$ on the background mesh computed as 
\begin{equation} \label{eq:uestimated}
\mathbf{\tilde{u}}(\mathbf{x})=\mathbf{u}^{n}(\mathbf{x})+\mathbf{rhs}^{n}(\mathbf{x})\Delta t,
\end{equation}
where $\Delta t$ is the size of the time step, the right-hand-side term $\mathbf{rhs}^{n}$ includes the convection, pressure gradient and diffusion terms computed from the quantities of previous time step $n$.  

The tangential force acting on the surface depends on the incoming velocity, the surface geometry and the complex near-wall turbulence.  Neither of the last two can be directly captured in the actuator type simulations by using the no-slip boundary conditions or the shear stress boundary conditions for wall models~\cite{cabot2000approximate, piomelli2002wall}. In the present actuator surface model, we assume that the tangential force is proportional to the local incoming velocity, and the effects of surface geometry and near-wall turbulence can be parameterized using a single parameter, the friction coefficient $c_{f}$. Based on this assumption, the tangential force acting on the surface per unit area is computed as 
\begin{equation}\label{eq:Ft_ASN}
\mathbf{f}_{\tau}(\mathbf{X})=\frac{1}{2}c_{f}U^2\mathbf{e}_\tau(\mathbf{X})
\end{equation}
where $c_{f}$ is calculated from the empirical relation proposed by F. Schultz-Grunow~\cite{schlichting2003boundary} for turbulent boundary layers with zero pressure gradient, which is expressed as follows:
\begin{equation}\label{eq:cf}
c_f=0.37(\log Re_x)^{-2.584},
\end{equation}
where $Re_x$ is the Reynolds number based on the incoming velocity and the distance from the upstream edge of the immersed body. It is noticed that this expression is invalid for the region with a large pressure gradient. For the present nacelle simulation, however, zero pressure gradient can be regarded as a reasonable assumption for the most part of the turbine nacelle. The framework presented in this work is applicable to more complex geometries if the corresponding distribution of $c_f$ is available from experiments or high fidelity simulations. 
In Eq.~(\ref{eq:Ft_ASN}), $U$ is the reference velocity of the incoming flow. In the present work, it is selected as the magnitude of the incoming streamwise velocity.  The direction of the tangential force $\mathbf{e}_\tau(\mathbf{X})$, on the other hand, is determined by the local tangential velocity relative to the velocity at the nacelle surface at points located at $h$ away from the wall, i.e., point B in Fig.~\ref{fig:nacellemodel}, as follows:
\begin{equation}\label{eq:ntang}
\mathbf{e}_\tau(\mathbf{X}) = \frac{\mathbf{u}\left(\mathbf{X}+h\mathbf{e}_n\left(\mathbf{X}\right)\right)}{|\mathbf{u}\left(\mathbf{X}+h\mathbf{e}_n\left(\mathbf{X}\right)\right)|}.
\end{equation}

After computing the forces on the actuator surface of the nacelle, the forces on the background mesh are distributed from the actuator surface mesh in the same way as in Eq.~(\ref{eq:force-dist}). 
It is noted that the forces are distributed over the closest five cells defined by the width of the employed smoothed cosine discrete delta function. This force distribution width will be very small if we have a sufficiently high grid resolution to resolve the boundary layer. For the grids using in the actuator type simulations, however, this force distribution width is so large that it may cause non-physical velocity field near the nacelle and affect the calculation of the relative incoming velocity for the blade. To remedy this problem, we simply let the relative incoming velocity ($\mathbf{V}_{rel}$ in Eq.~(\ref{eq:Vrel})) of the two closest radial locations to the nacelle surface be equal to that of the third closest location away from the nacelle. 
\section{Flow solver}
\label{sec:flowsolver}
The governing equations are the three-dimensional, unsteady, filtered continuity and Navier-Stokes equations in non-orthogonal, generalized, curvilinear coordinates, which read in compact tensor notation (repeated indices imply summation) as follows ($i, j = 1, 2, 3$):
\begin{equation}
J\frac{\partial U^{i}}{\partial \xi^{i}}=0,
\label{eq:continuity}
\end{equation}
\begin{align}
\frac{1}{J}\frac{\partial U^{i}}{\partial t}=& \frac{\xi _{l}^{i}}{J}\left( -%
\frac{\partial }{\partial \xi^{j}}({U^{j}u_{l}})+\frac{\mu}{\rho}%
\frac{\partial }{\partial \xi^{j}}\left(  \frac{g^{jk}}{J}\frac{%
\partial u_{l}}{\partial \xi^{k}}\right) -\frac{1}{\rho}\frac{\partial }{\partial \xi^{j}}(\frac{%
\xi _{l}^{j}p}{J})-\frac{1}{\rho}\frac{\partial \tau _{lj}}{\partial
\xi^{j}}+f_{l}\right) ,
\label{eq:momentum}
\end{align}%
where $x_i$ and $\xi^i$ are the Cartesian and curvilinear coordinates, respectively, $\xi _{l}^{i}={\partial \xi^{i}}/{\partial x_{l}}$ are the transformation metrics, $J$ is the Jacobian of the geometric transformation, $u_{i}$ is the $i^{th}$ component of the velocity vector in Cartesian coordinates, $U^{i}$=${(\xi _{m}^{i}/J)u_{m}}$ is the contravariant volume flux, $g^{jk}=\xi _{l}^{j}\xi _{l}^{k}$ are the components of the contravariant metric tensor, $\rho $ is the density, $\mu $ is the dynamic viscosity, $p$ is the pressure,  $f_l (l=1,2,3)$ are the body forces introduced by the turbine blade and nacelle computed using the actuator surface models presented in Sec.~\ref{sec:ASM}, and $\tau_{ij}$ represents the anisotropic part of the subgrid-scale stress tensor, which is modelled by the dynamic eddy viscosity subgrid-scale stress model~\cite{germano1991dynamic}. \textcolor[rgb]{0,0,0}{The governing equations are discretized in space using a second-order accurate central differencing scheme, and integrated in time using the fractional step method~\cite{ge2007numerical}. An algebraic multigrid acceleration along with GMRES solver is used to solve the pressure Poisson equation. A matrix-free Newton-Krylov method is used for solving the discretized momentum equation. }
\section{Test cases}
\label{sec: testcases}
In this section, we evaluate the performance of the proposed actuator surface models for turbine blade and nacelle. The nacelle body case in Sec.~\ref{sec: nacelle} is to evaluate the capability of the actuator surface model for nacelle. The MEXICO (Model experiments in Controlled Conditions) turbine~\cite{schepers2007model, shen2012actuator, nilsson2015validation} case in Sec.~\ref{sec: MEXICO} is to show the capability of the actuator surface models in predicting the tip vortices and velocity distribution in the near wake. The capability of the new actuator surface models on the wake meandering prediction is demonstrated by the hydrokinetic turbine case in Sec.~\ref{sec: mhkturbine}.  In turbine simulations the proposed actuator surface model is abbreviated as ASB-ASN when both the actuator surface blade and actuator surface nacelle models are employed, and ASB when only the blade model is employed. \textcolor[rgb]{0,0,0}{The conventional actuator line simulations are also carried out for comparison. The actuator line model for turbine blade is abbreviated as ALB. The actuator line model for blades with the actuator surface model for nacelle is abbreviated as ALB-ASN.} Throughout this paper, $x$, $y$ and $z$ represent the streamwise, spanwise and vertical directions, respectively. 
\subsection{Flow over periodically placed nacelles}
\label{sec: nacelle}
\begin{figure}
  \centerline{\includegraphics[width=2.5in]{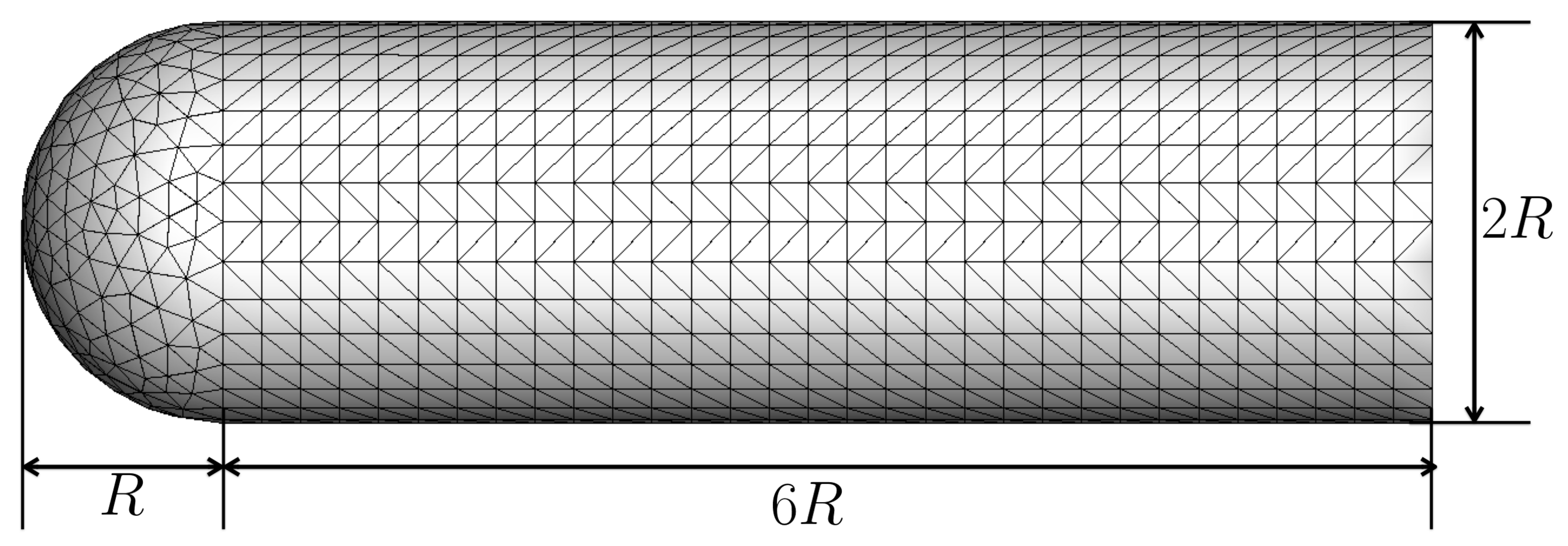}}
  \caption{Schematic of the nacelle geometry with the triangular surface mesh. The flow is from left to right. }
\label{fig:schematic_nacelle}
\end{figure}

In this section, the actuator surface model for nacelle is applied to simulate the flow over periodically placed nacelles to mimic wind farm scenarios. The nacelle body is composed of a hemisphere facing upstream and a circular cylinder in its downstream as shown in Fig.~\ref{fig:schematic_nacelle}. Free-slip boundary condition is applied in the crosswise directions. Periodic boundary condition is employed in the streamwise direction. The Reynolds number based on the cylinder radius $R$ and the freestream velocity is 1000. 
The size of the computational domain is $L_x \times L_y \times L_z = 30R \times 20R \times 20R$. 
The downstream end of the nacelle body is located at the origin of the streamwise coordinate. 
Three grids are employed with the number of grid nodes $N_x \times N_y \times N_z = 502 \times 348 \times 348$,  $N_x \times N_y \times N_z = 115 \times 151 \times 151$ and $N_x \times N_y \times N_z = 153 \times 80 \times 80$, grid spacings near the nacelle $\Delta x = R/8, R/3.75$ and $R/2.5$, $\Delta y = \Delta z = R/33, R/7.5$ and $R/5$, and the corresponding time steps $0.01R/U$, $0.1R/U$ and $0.1R/U$ for the fine, medium and coarse grids, respectively. 
 Large-eddy simulation with the dynamic subgrid-scale model~\cite{germano1991dynamic} is employed for the flow simulation~\cite{kang2011high}. \textcolor[rgb]{0,0,0}{On the coarse grid, two actuator type large-eddy simulations are carried out: one uses the proposed actuator surface model for nacelle; the other one represents the nacelle by a permeable disk located at the downstream end of the nacelle. In the permeable disk simulation, the mean drag force computed in the coarse grid actuator surface simulation is uniformly distributed on the disk, which gives a drag coefficient of $C_D = 0.48$ defined using the freestream velocity and the area $\pi R^2$. On the medium grid, only the actuator surface simulation is carried out. }  On the fine grid, the geometry-resolving, wall-resolved large-eddy simulation is carried out with the sharp-interface immersed boundary method developed by Ge et al.~\cite{ge2007numerical} for resolving the boundary layer and near wall turbulence, from which the computed results is employed to gauge the performance of the proposed actuator surface model. \textcolor[rgb]{0,0,0}{In both actuator surface and immersed boundary simulations, the number of surface triangular cells is 2652. In the permeable disk simulation, the number of surface triangular cells is 234. }
%
%It is noticed that two types of geometry-resolving large-eddy simulations have been mentioned: the geometry-resolving, wall-modeled large-eddy simulation as employed in~\cite{kang2014onset} and the geometry-resolving, wall-resolved large-eddy simulation employed in the fine grid nacelle simulation. 
%
%

%The contours of the time-averaged streamwise velocity computed from the coarse grid actuator surface simulation is compared with the fine grid wall-resolved LES in Fig.~\ref{fig:contour_uavg_nacelle}. As seen, the shape of the wake and the velocity at far wake locations predicted by the actuator surface model show overall good agreement with the wall-resolved LES results. While in the near wake, the recirculation bubble computed from the actuator surface model is somewhat longer for that computed from the actuator surface model. 
%
%\begin{figure}
%  \centerline{\includegraphics[width=\textwidth, height=\textheight,keepaspectratio]{contour_uavg_nacelle}}
%  \caption{Contours of time-averaged streamwise velocity of the nacelle case. Left: wall-resolved LES; Right: actuator surface simulation. The white lines show the zero contour level of $\langle u \rangle/U$. }
%\label{fig:contour_uavg_nacelle}
%\end{figure}
%

\begin{figure}
  \centerline{\includegraphics[width=\textwidth]{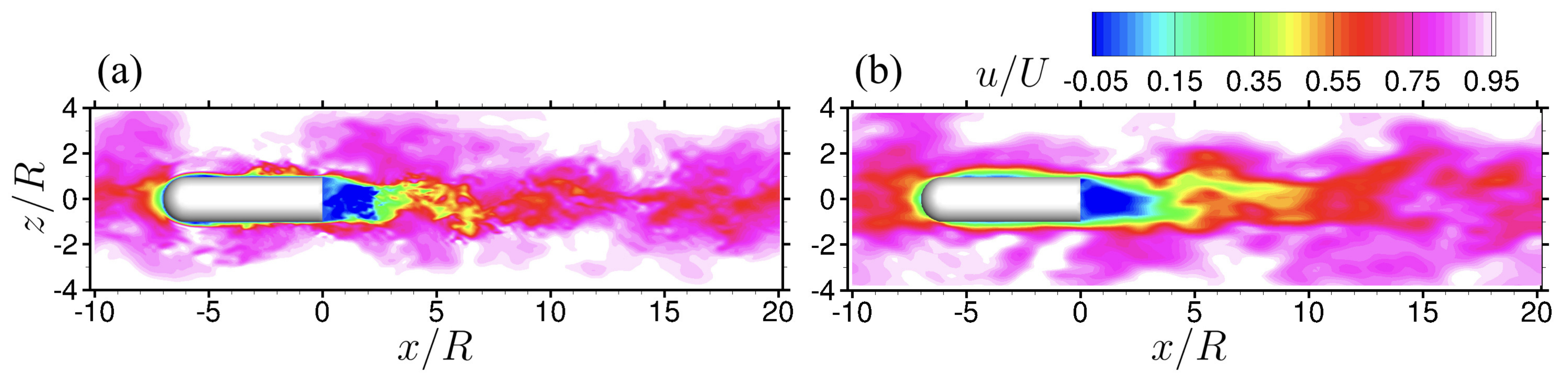}}
  \caption{Contours of instantaneous streamwise velocity of the nacelle case. (a): wall-resolved LES; (b): actuator surface simulation (coarse grid).} 
\label{fig:uinst_nacelle}
\end{figure}

\textcolor[rgb]{0,0,0}{The contours of the instantaneous streamwise velocity computed from the coarse grid actuator surface simulation is compared with the wall-resolved LES in Fig.~\ref{fig:uinst_nacelle}. As seen, the large-scale flow structure in the nacelle wake is well predicted by the actuator surface model on a very coarse grid.}

\begin{figure}
  \centerline{\includegraphics[width=0.8\textwidth, height=\textheight,keepaspectratio]{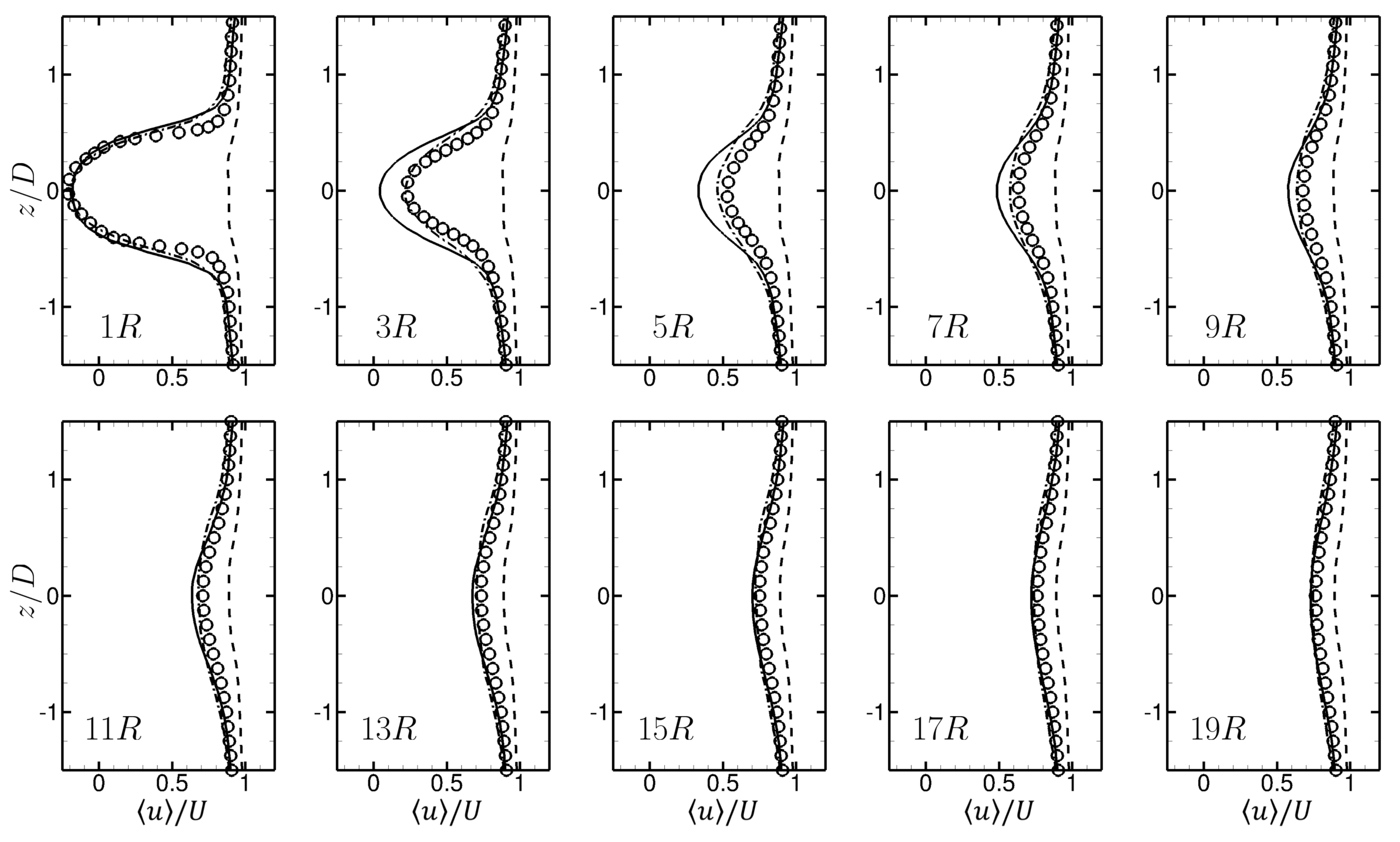}}
  \caption{Vertical profiles of time-averaged streamwise velocity $\left<u\right>$ at different downstream locations for the nacelle case. Symbols: wall-resolved LES; Solid lines: actuator surface simulation on the coarse grid; Dash-dot lines: actuator surface simulation on the medium grid; Dashed lines: permeable disk simulation on the coarse grid.}
\label{fig:u_nacelle}
\end{figure}

The mean velocity profiles computed from the actuator type models are compared with that from the wall-resolved LES in Fig.~\ref{fig:u_nacelle}. As seen, the velocity deficit computed from actuator surface model agrees well with that from the wall-resolved LES at $1R$ downstream. At $3R$, $5R$ and $7R$ downstream locations, however, the velocity deficits computed from the actuator surface model on the coarse grid are significantly larger. For further downstream locations, on the other hand, the velocity deficits predicted by the actuator surface model show good agreement with that computed by the wall-resolved LES. \textcolor[rgb]{0,0,0}{The velocity deficits computed by the actuator surface model on the medium grid show good agreements with that from the wall-resolved LES at all downstream locations. The permeable disk model significantly underpredicts the velocity deficits at all downstream locations, although the drag force with the same magnitude is applied on the permeable disk.}

The turbulence kinetic energy  profiles computed from the actuator type models are compared with that from the wall-resolved LES in Fig.~\ref{fig:u_nacelle}. At $1R$ downstream location, the actuator surface model gives acceptable predictions on both coarse and medium grids. At $3R$ downstream location, the actuator surface model significantly underpredicts the turbulence kinetic energy on the coarse grid. On the medium grid, on the other hand, a significant improvement is observed. At $7R$ to further downstream locations, the turbulence kinetic energy profiles computed on coarse and medium grids are similar with each other and are larger that that from the wall-resolved LES. It is also noticed that the peaks of turbulence kinetic energy predicted by the actuator surface simulations are different from that predicted by wall-resolved LES at $3R$, $5R$ and $7R$ downstream locations. The actuator disk model, on the other hand, predicts nearly negligible turbulence kinetic energy at all downstream locations. 
\begin{figure}
  \centerline{\includegraphics[width=0.8\textwidth, height=\textheight,keepaspectratio]{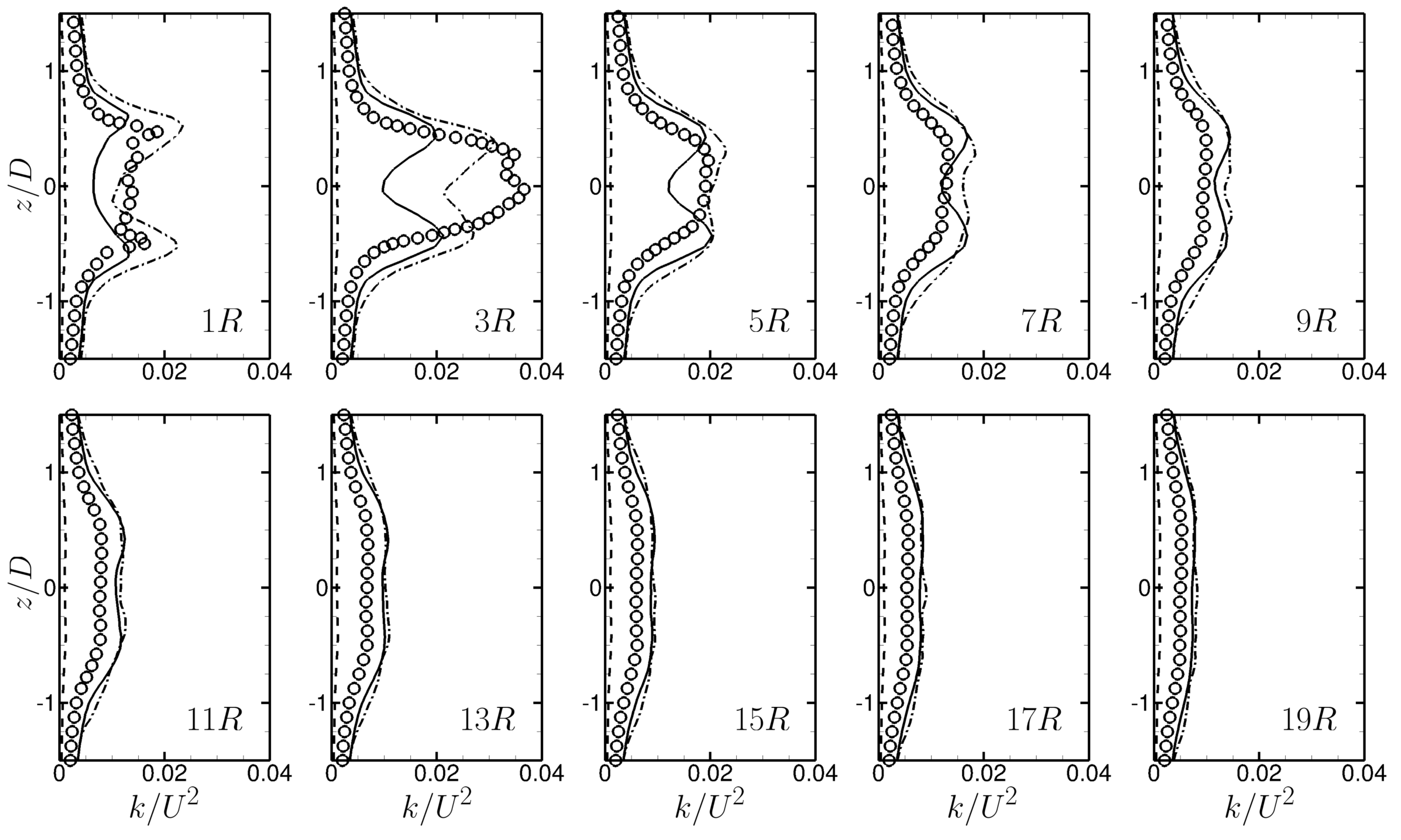}}
  \caption{Vertical profiles of turbulence kinetic energy $k$ at different downstream locations for the nacelle case. Symbols: wall-resolved LES; Solid lines: actuator surface simulation on the coarse grid; Dash-dot lines: actuator surface simulation on the medium grid; Dashed lines: permeable disk simulation on the coarse grid.}
\label{fig:k_nacelle}
\end{figure}

\begin{figure}
  \centerline{\includegraphics[width=0.8\textwidth, height=\textheight,keepaspectratio]{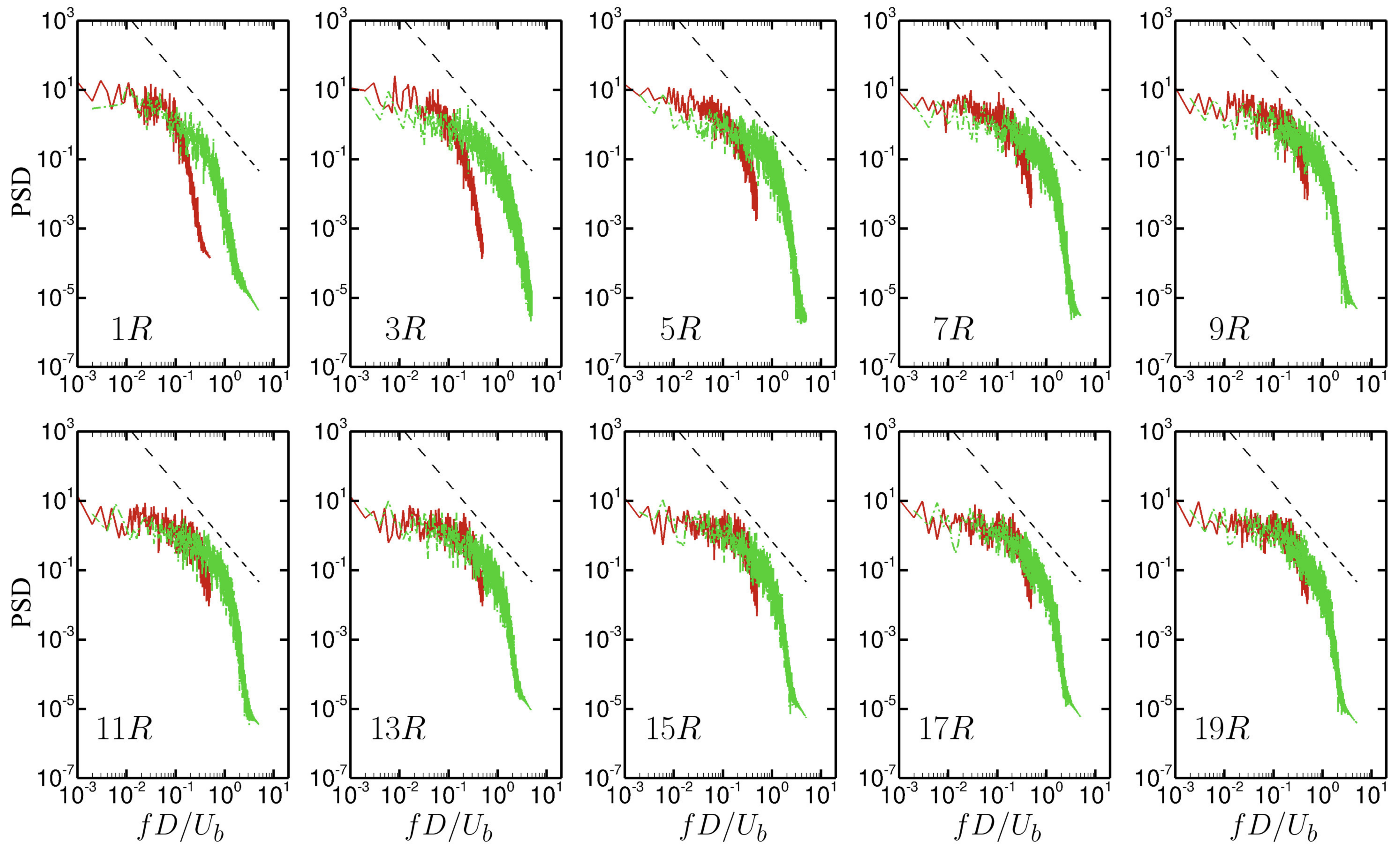}}
  \caption{Power spectral density (PSD) of streamwise velocity fluctuations at different downstream locations for the nacelle case. Red solid lines: actuator surface simulation on the coarse; Green dash-dot lines: wall-resolved LES. The black dashed line shows the slope of $-5/3$.}
\label{fig:psd_nacelle}
\end{figure}

To further evaluate the performance of the actuator surface model, the power spectral density (PSD) predicted by the actuator surface simulation is compared with that from the wall-resolved LES at different downstream locations in Fig.~\ref{fig:psd_nacelle}. It is seen the low frequency, energy-containing part of the PSD predicted by the actuator surface model shows an overall good agreement with the wall-resolved LES. Both the actuator surface and wall-resolved simulations predict the inertial range of the PSD with the slope of $-5/3$. \textcolor[rgb]{0,0,0}{However, the inertial range predicted in the actuator surface simulation is narrower than that from the wall-resolved LES at $1R$, $3R$ and $5R$ downstream locations. This implies that the low spatial resolution is the main reason for the discrepancy observed in the turbulence kinetic energy profiles in the near wake locations (Fig.~\ref{fig:k_nacelle}), where the small scale energetic turbulent flow structures are not resolved by the grid.} At far wake locations, both energy-containing part and inertial part of the PSD predicted by the actuator surface model agree well with that computed from the wall-resolved LES, even though the very high frequency, dissipation range is not resolved in the actuator surface simulation.  
\subsection{Flow over the MEXICO turbine}
\label{sec: MEXICO}
In this section, we simulate the flow over the MEXICO turbine to evaluate the performance of the proposed actuator surface models on predicting the velocity and tip vortices at turbine near wake locations. The size of the computational domain is $L_x \times L_y \times L_z = 4.2D \times 4D \times 4D$. The number of grid nodes is $N_x \times N_y \times N_z = 673 \times 502 \times 502$. The mesh is uniform across the turbine and in the streamwise direction with grid spacing $D/160$, and is stretched in the spanwise and vertical directions at regions away from the turbine. The size of time step is $T/750$, where $T$ is rotational period of the turbine rotor. 
\textcolor[rgb]{0,0,0}{In the actuator surface simulations, the number of triangular cells for actuator blades and nacelle is 1461 and 5464, respectively. In the actuator line simulation, the number of segments is 141.} 
Free-slip boundary condition is applied at the boundaries in the spanwise and vertical directions. The flow at the inlet is uniform. Three cases with three different tip speed ratios ($\text{TSR}=\Omega R/U$ where $\Omega$ is the rotational speed of the rotor, $R$ is the radius of the rotor and $U$ is the incoming wind speed), i.e. 4.2, 6.7 and 10, are carried out. In the experiment, different tip speed ratios were realized by adjusting the incoming wind speed. In the LES, on the other hand, different rotational speeds are employed for different tip speed ratios. The Reynolds number based on the incoming wind speed and the rotor diameter is $Re=6.9\times10^6$  the same for the cases with different tip speed ratios. \textcolor[rgb]{0,0,0}{Both ASB-ASN and ALB simulations are carried out. In the ASB-ASN simulation, the nacelle is simply represented by a circular cylinder with diameter $0.2D$ and length $0.2D$.}

\begin{figure}
  \centerline{\includegraphics[width=0.8\textwidth, height=\textheight,keepaspectratio]{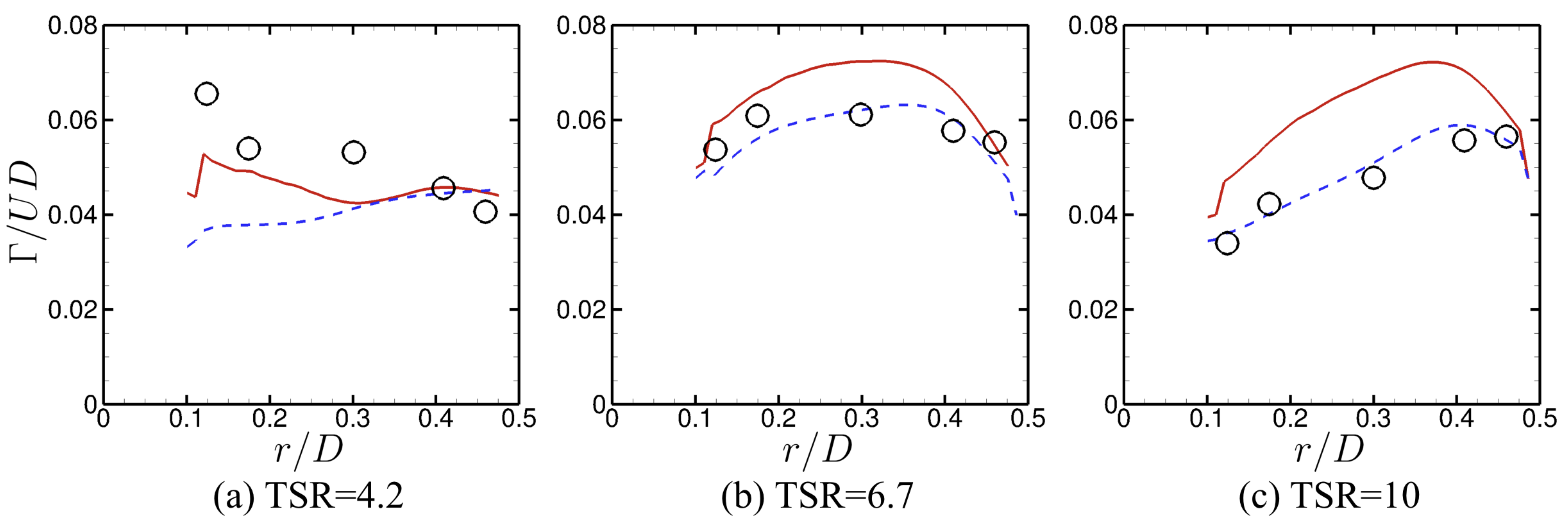}}
  \caption{Bound circulation distributions along blade radial direction for (a) TSR=4.2, (b) TSR=6.7 and (c) TSR=10. Circles: measurements; Red solid line: ASB-ASN; Blue dashed line: ALB.  }
\label{fig:boundcirculation_MEXICO}
\end{figure}

Before presenting the computed results in the turbine near wake, the bound circulations along the blade predicted by the actuator surface model and the conventional actuator line model are compared with measurements in Fig.~\ref{fig:boundcirculation_MEXICO}. The bound circulation is computed by $\Gamma = \frac{L}{\rho |\mathbf {V}_{rel}|}$. As seen, the location with maximum bound circulation is located near the blade root for TSR=4.2, in the middle for TSR=6.7, while close to the tip for TSR=10.  Such dependence of the bound circulation distributions on the values of TSR is well captured by the proposed actuator surface model. The amplitudes of the predicted bound circulation also show an acceptable agreement with the measurements except for the TSR=10 case, where the discrepancies between the computed values and measured ones are somewhat larger than the other two cases. \textcolor[rgb]{0,0,0}{The actuator line model predicts with good accuracy the bound circulation for the TSR=6.7 and 10 cases, while underpredicts the bound circulation for the TSR=4.2 case.} 

\begin{figure}
  \centerline{\includegraphics[width=0.8\textwidth, height=\textheight,keepaspectratio]{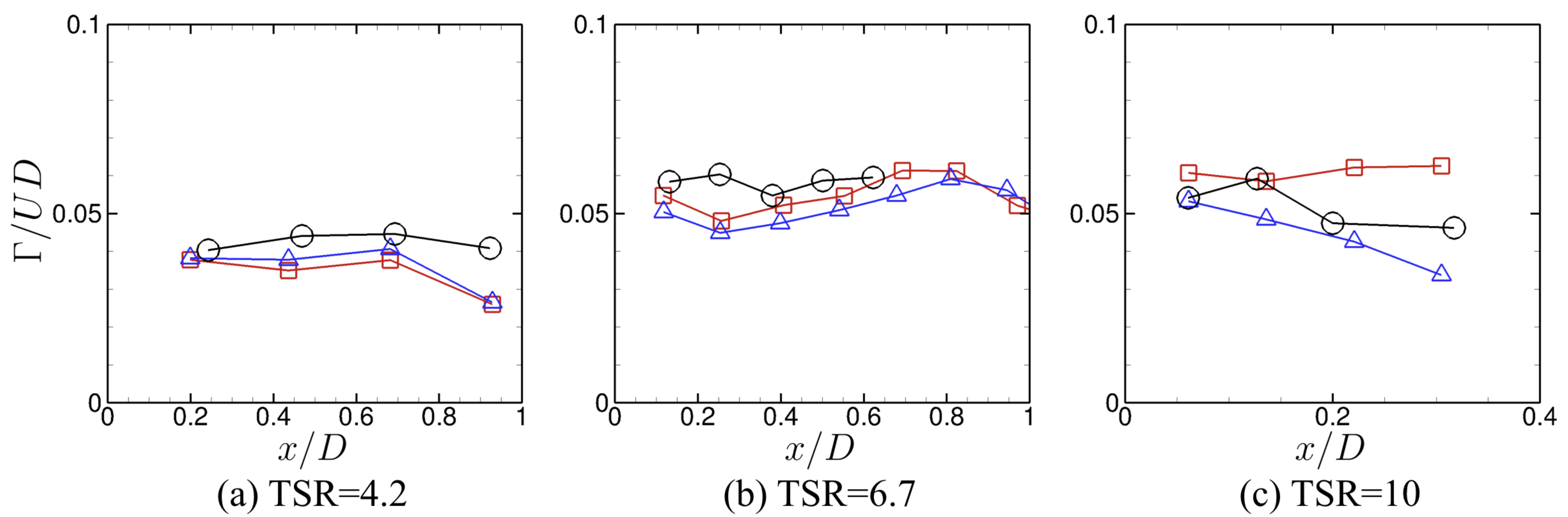}}
  \caption{Circulation of tip vortices at different downwind locations for (a) TSR=4.2, (b) TSR=6.7 and (c) TSR=10. Black line with circles: measurements; Red line with squares: ASB-ASN; Blue line with triangles: ALB.}
\label{fig:tipcirculation_MEXICO}
\end{figure}

The capability of the proposed actuator surface model in predicting the tip vortices is examined in Fig.~\ref{fig:tipcirculation_MEXICO}, in which we show the comparison of the computed tip vortex circulations with the measurements for the three cases with different tip speed ratios. To calculate the circulation of tip vortices, the center of a tip vortex core is selected as the position with the maximum vorticity magnitude. The circulation is computed by integrating the velocity along a circle centered at the vortex core center with radius of $0.05D$. The computed tip vortex circulations show overall good agreements with the measurements for the actuator surface model. Some discrepancies are observed for the TSR=10 case for the third and the fourth tip vortices downwind from the rotor for the actuator surface model. While the circulation of tip vortices predicted by the actuator line model agrees well the measurements for the TSR=10 case. From the comparison of the tip vortex circulations among the three TSR cases, it is interesting to notice that they are in the same order when normalized using $U$ and $D$, although the dimensional circulations are significantly different for the three cases. 

\begin{figure}
  \centerline{\includegraphics[width=0.8\textwidth]{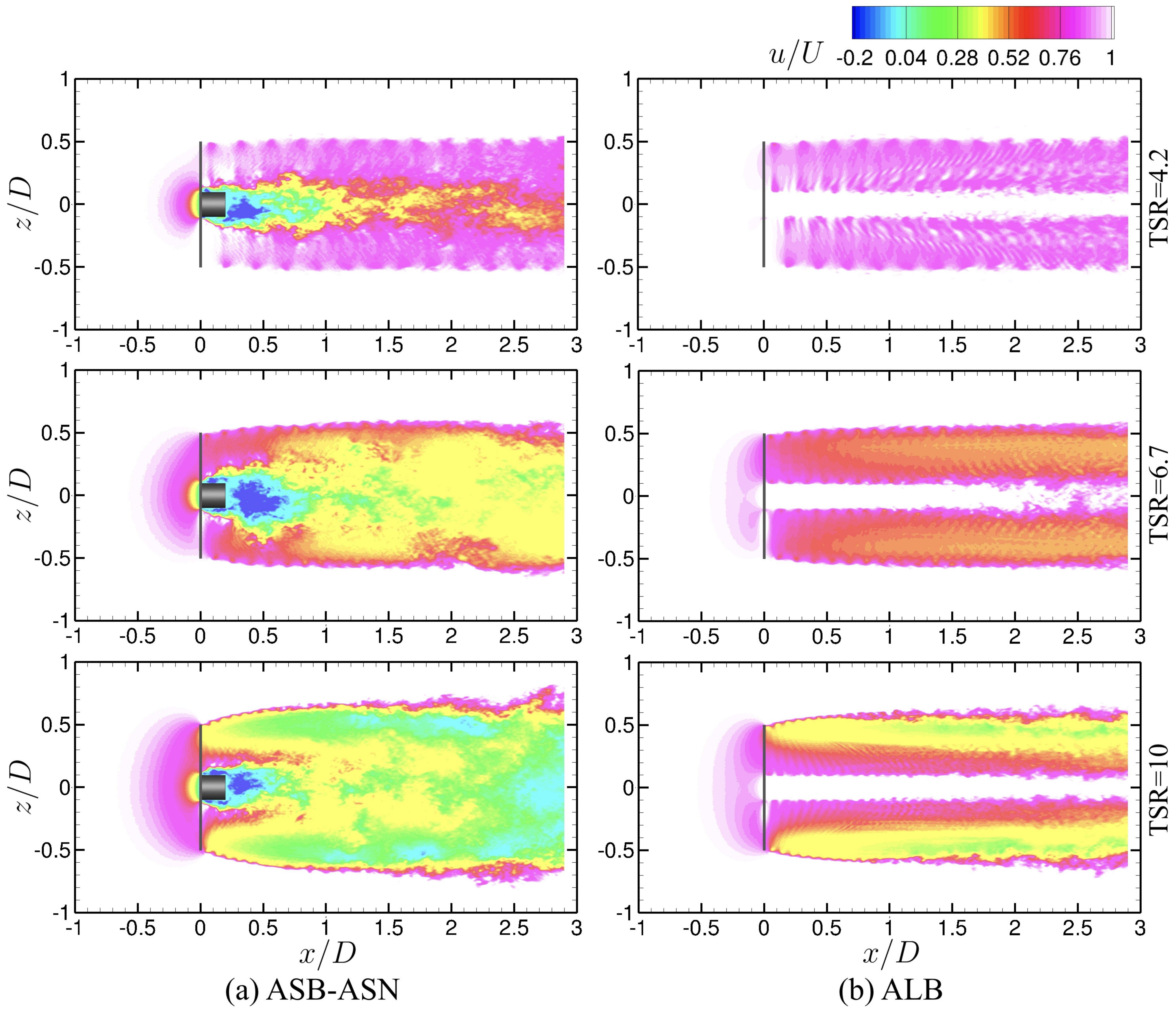}}
  \caption{Contours of instantaneous streamwise velocity of the MEXICO turbine case. (a): ASB-ASN; (b): ALB.} 
\label{fig:uinst_mexico}
\end{figure}

\textcolor[rgb]{0,0,0}{The contours of the instantaneous streamwise velocity are shown in Fig.~\ref{fig:uinst_mexico} for different tip-speed ratios for both ASB-ASN and ALB simulations. As seen, a jet exists at the center in the ALB results. On the other hand, a wake from the nacelle forms in the ASB-ASN results. This nacelle wake substantially interacts with the outer part of the wake and causes intensive velocity fluctuations in the turbine wake. Although significantly differences in turbulence statistics between ASB-ASN and ALB results are expected, this work will only focus on the mean velocity profiles for which the wind tunnel measurements are available for validation purpose. }

\begin{figure}
  \centerline{\includegraphics[width=0.8\textwidth, height=\textheight,keepaspectratio]{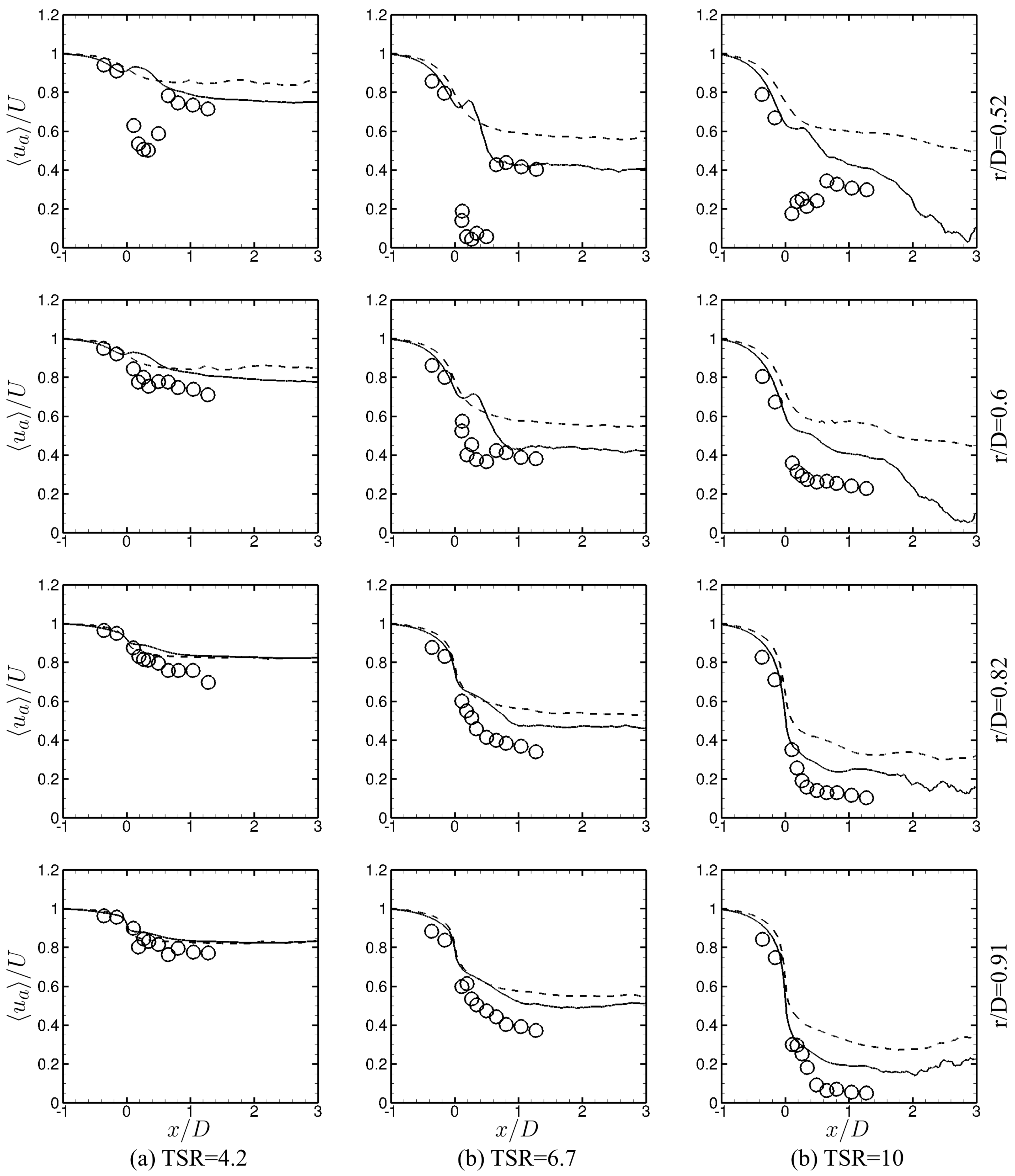}}
  \caption{Axial velocity profiles for the flow past the MEXICO turbine at different radial locations for (a) TSR=4.2, (b) TSR=6.7 and (c) TSR=10. Circles: measurements; Solid line: ASB-ASN; Dashed line: ALB. }
\label{fig:MEXICO_uaprofile}
\end{figure}

\begin{figure}
  \centerline{\includegraphics[width=0.8\textwidth, height=\textheight,keepaspectratio]{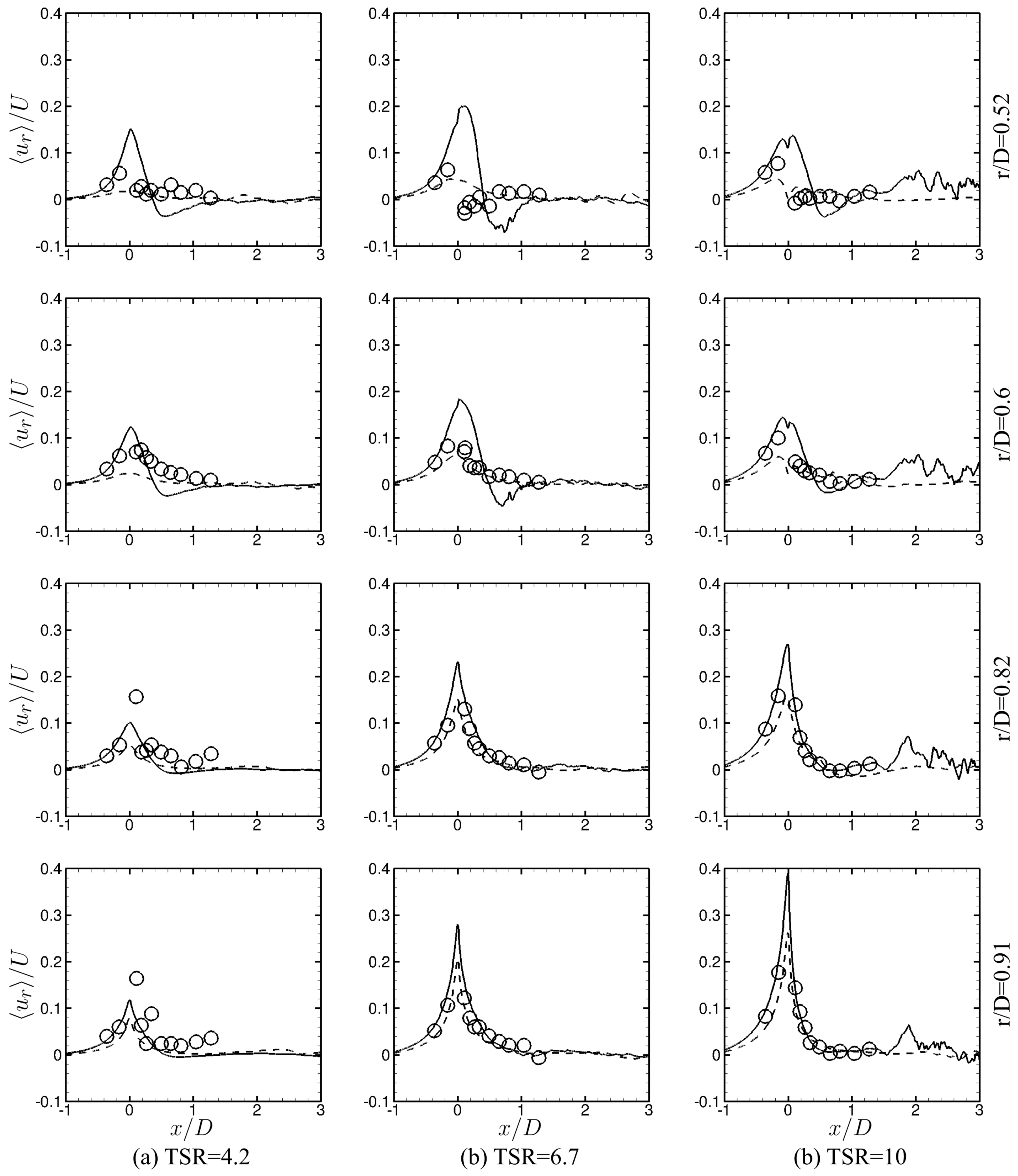}}
  \caption{Radial velocity profiles for the flow past the MEXICO turbine at different radial locations for (a) TSR=4.2, (b) TSR=6.7 and (c) TSR=10. Circles: measurements; Solid line: ASB-ASN; Dashed line: ALB. }
\label{fig:MEXICO_urprofile}
\end{figure}

The computed axial ($\langle u_a \rangle$) and radial ($\langle u_r \rangle$) velocities are compared with measurements at different radial locations in Fig.~\ref{fig:MEXICO_uaprofile} and Fig.~\ref{fig:MEXICO_urprofile}, respectively. As seen, the radial velocity $\langle u_r \rangle$ increases as the flow approaches turbine, and decreases rapidly in the immediate downwind of the turbine for all the four considered radial locations for all the three cases. It is also seen that such increase of the radial velocity is somewhat larger when TSR is higher. From these comparisons, we can see that the developed actuator surface model can accurately predict the downwind variations of radial velocity. For the axial velocity, a sudden decrease and increase in the near wake of the turbine are observed at $r/D=0.52$ for all the three tip-speed ratios for the measured data. This is, however, because of the corrupted PIV photos at these locations caused by the light reflected on the turbine hub as noticed in the paper by Nilsson et al.~\cite{nilsson2015validation}. The experimental data from the corrupted PIV images will be ignored for the comparisons. It is seen that overall good agreements with measurements are obtained for the axial velocities although some discrepancies are observed for the TSR=10 case. \textcolor[rgb]{0,0,0}{The actuator line model, on the other hand, underpredicts the velocity deficits for the TSR=6.7 and 10 cases. }
\subsection{Flow over a hydrokinetic turbine}
\label{sec: mhkturbine}
The capability of the developed actuator surface models in predicting the wake meandering at far wake locations is assessed by simulating the turbulent flow over the axial-flow hydrokinetic turbine at Saint Anthony Falls Laboratory, University of Minnesota~\cite{chamorro2013interaction}. Kang, Yang and Sotiropoulos~\cite{kang2014onset} carried out the geometry-resolving, actuator disk and actuator line simulations of this turbine. It was found that the conventional actuator line model without a nacelle model underpredicts the turbulence intensity in the far wake because it cannot accurately capture the inner-outer wake interaction and thus the wake meandering in the far wake. 

The computational domain is $L_x \times L_y \times L_z= 16D \times 5.5D \times 2.3D$, where $D$=0.5 m is the rotor diameter. Two different grids are employed: a coarse grid with the grid spacing $D/50$ and a fine grid with grid spacing  $D/100$ around the turbine.  The numbers of grid nodes are $N_x \times N_y \times N_z= 802 \times 277 \times 116$ and $556 \times 341 \times 201$ with the corresponding time step $0.001D/U$ and $0.002D/U$ for the fine and coarse grids, respectively. 
\textcolor[rgb]{0,0,0}{For the coarse grid actuator surface cases, the number of triangular cells for actuator blades and nacelle is 912 and 1108, respectively. In the coarse grid actuator line cases, the number of segments is 78. For the find grid case, the number of triangular cells for actuator blades and nacelle is 1983 and 1636, respectively.} 
On the coarse grid, four simulations are carried out using ASB-ASN, ASB, ALB and ALB-ASN models, respectively. On the fine grid only the ASB-ASN simulation is carried out. The smoothed discrete delta function is employed for the force distribution. In the coarse grid simulation, \textcolor[rgb]{0,0,0} {the $h_x$, $h_y$ and $h_z$ in Eq.~(\ref{eq:phi_4sc}) are} equal to the grid spacing $D/50$. In order to ensure the same width of force distribution for the two different grids, \textcolor[rgb]{0,0,0} {the $h_x$, $h_y$ and $h_z$} are set as $D/50$ in the fine grid simulation. However, no significant difference is observed when $h_x$, $h_y$ and $h_z$ are equal to the grid spacing ($D/100$) for the fine grid simulation. 
The Reynolds number based on the bulk velocity and the rotor diameter is $2\times 10^5$. In the simulations, all the cases are first carried out until the total kinetic energy reaches a quasi-steady state, and subsequently the flowfields are averaged for $370T$ and $185T$ for the coarse grid and fine grid simulations, respectively, where $T=2\pi/\Omega$ is the rotor revolution period. The rotational speed of the rotor is $9.28$ $rad/s$, which gives a tip-speed ratio of 5.8 based on the bulk velocity (0.4 m/s).

\begin{figure}
  \centerline{\includegraphics[width=0.8\textwidth, height=\textheight,keepaspectratio]{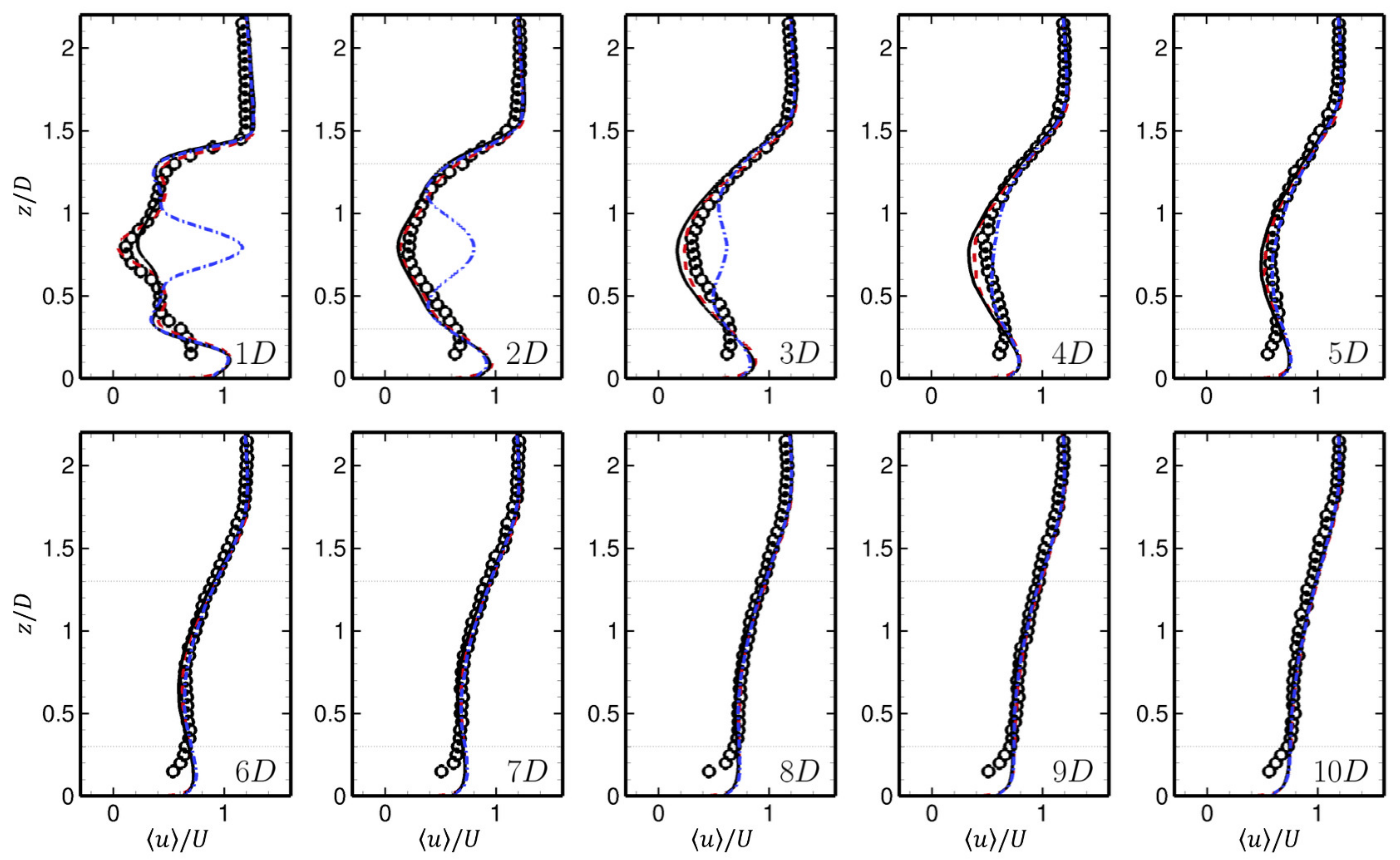}}
  \caption{Vertical profiles of the time-averaged streamwise velocity $\left<u\right>$ at different downstream locations for the hydrokinetic turbine case. If not indicated, the computed results are from cases with grid spacing $D/50$. Symbols: measurements from~\cite{chamorro2013interaction}; Black solid line: ASB-ASN; Red dashed line: ASB-ASN with grid spacing $D/100$; Blue dash-dot line: ASB.}
\label{fig:uavgprof_mhk_ASB}
\end{figure}

\begin{figure}
  \centerline{\includegraphics[width=0.8\textwidth, height=\textheight,keepaspectratio]{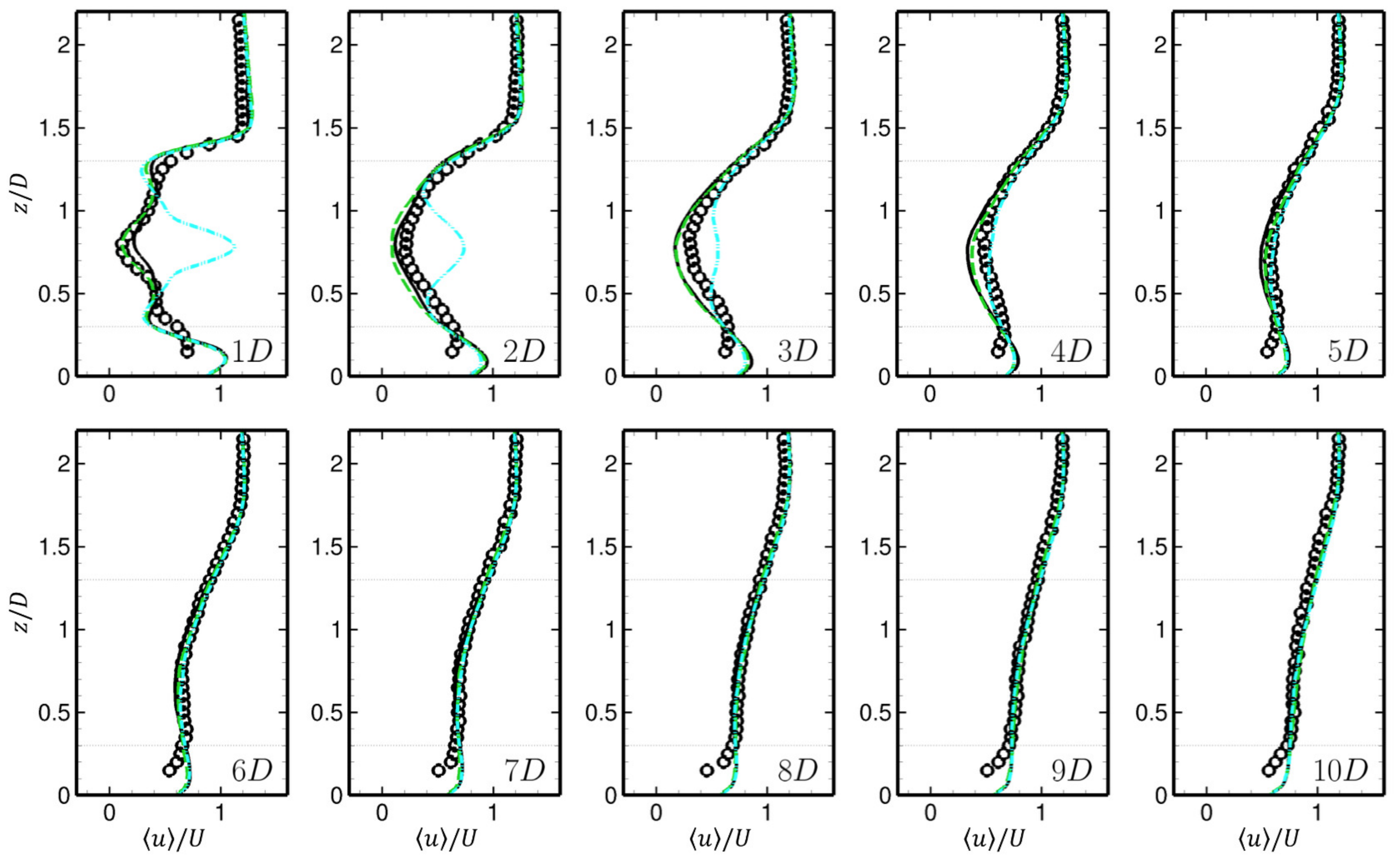}}
  \caption{Vertical profiles of the time-averaged streamwise velocity $\left<u\right>$ at different downstream locations for the hydrokinetic turbine case. If not indicated, the computed results are from cases with grid spacing $D/50$. Symbols: measurements from~\cite{chamorro2013interaction}; Black solid line: ASB-ASN; Green long-dash line: ALB-ASN; Cyan dash-dot-dot line: ALB.}
\label{fig:uavgprof_mhk_ASB-ALB}
\end{figure}

We first examine the vertical profiles of the streamwise velocity and turbulence kinetic energy at different downstream locations. Comparison of the computed streamwise velocity profiles with the measurements is shown in Figs.~\ref{fig:uavgprof_mhk_ASB} and \ref{fig:uavgprof_mhk_ASB-ALB}. As expected, the ASB and ALB simulations without a nacelle model predict higher streamwise velocity near the hub at near wake locations ($1D$, $2D$ and $3D$). With the actuator surface model for nacelle, significant improvements are obtained on predicting the streamwise velocity near hub in the near wake for both actuator surface and actuator line blade models.  For further downstream locations from $4D$ to $10D$, the predicted streamwise velocities, on the other hand, are nearly the same for all the  simulations.

\begin{figure}
  \centerline{\includegraphics[width=0.8\textwidth, height=\textheight,keepaspectratio]{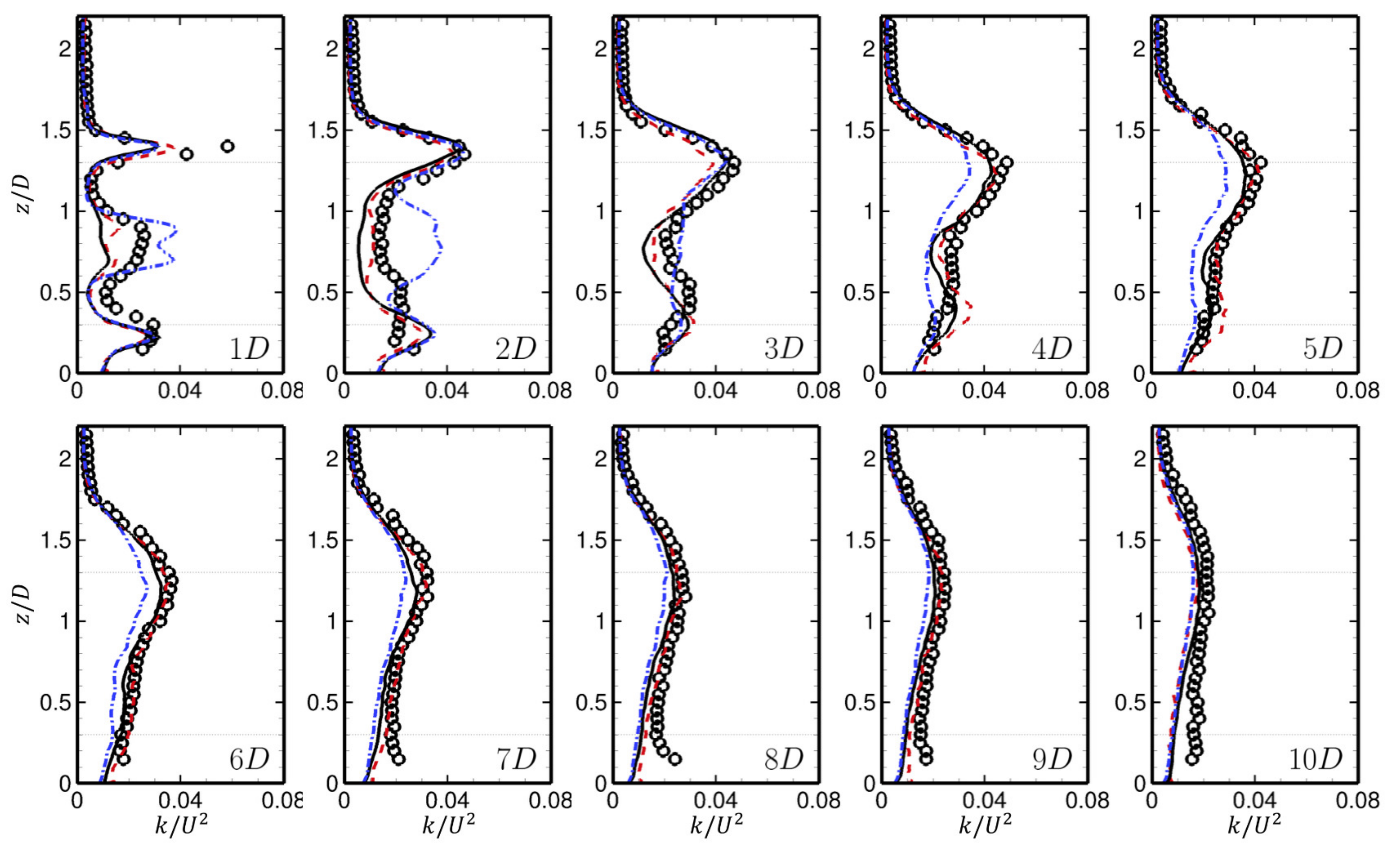}}
  \caption{Vertical profiles of the turbulence kinetic energy $k$ at different downstream locations for the hydrokinetic turbine case. If not indicated, the computed results are from cases with grid spacing $D/50$. Symbols: measurements from~\cite{chamorro2013interaction}; Black solid line: ASB-ASN; Red dashed line: ASB-ASN with grid spacing $D/100$; Blue dash-dot line: ASB.}
\label{fig:tkeprof_mhk_ASB}
\end{figure}

\begin{figure}
  \centerline{\includegraphics[width=0.8\textwidth, height=\textheight,keepaspectratio]{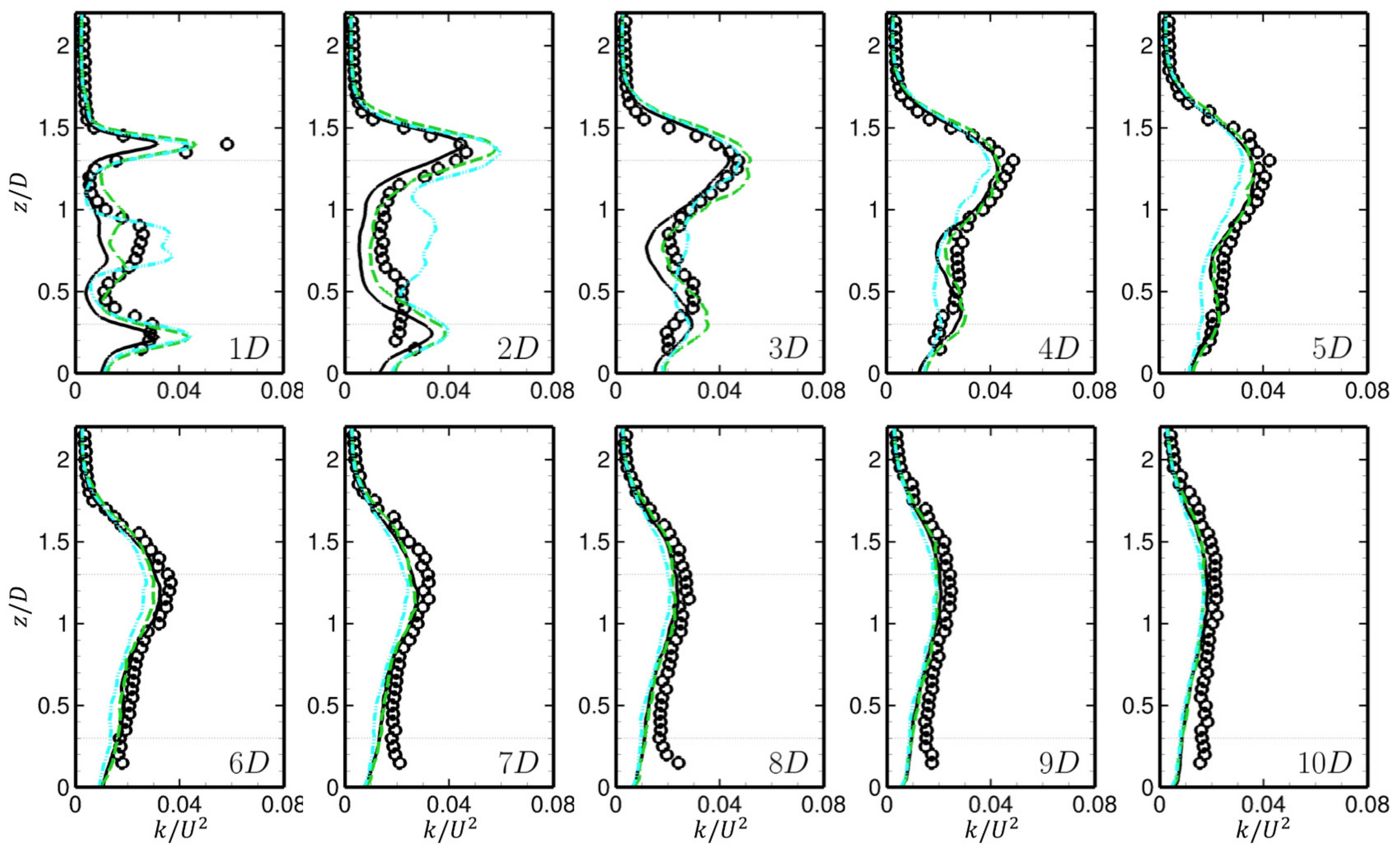}}
  \caption{Vertical profiles of the turbulence kinetic energy $k$ at different downstream locations for the hydrokinetic turbine case. If not indicated, the computed results are from cases with grid spacing $D/50$. Symbols: measurements from~\cite{chamorro2013interaction}; Black solid line: ASB-ASN; Cyan dash-dot-dot line: ALB; Green long-dash line: ALB-ASN.}
\label{fig:tkeprof_mhk_ASB-ALB}
\end{figure}

Turbulence kinetic energy \textcolor[rgb]{0,0,0}{at turbine's far wake locations} can be considered as the footprint of wake meandering. In order to evaluate the capability of the developed actuator surface model in capturing wake meandering in turbine's far wake, we compare the computed turbulent kinetic energy ($k$) profiles with measurements in Figs.~\ref{fig:tkeprof_mhk_ASB} and ~\ref{fig:tkeprof_mhk_ASB-ALB}. At $1D$ turbine downstream, in the region near the bottom tip, the computed $k$ agrees well with the measurements for both ASB and ASB-ASN simulations.  Near the top tip region, the computed $k$ from all the simulations is lower than that from the experiment, while the $k$ computed from the ASB-ASN on the fine grid, the ALB and ALB-ASN simulations shows a better agreement. The largest differences among different simulation are observed in the region near the hub: the $k$ from the simulation with the actuator surface nacelle model is lower than the measurements; the $k$ from the simulations without the nacelle model, on the other hand, is larger than the measurements. At $2D$ turbine downstream, the $k$ profiles computed from all the ASB-ASN and ASB simulations are in good agreement with measurements in the region near the top tip, which are overpredicted by the ALB-ASN and ALB models. The $k$ profiles predicted by the simulations with the actuator surface nacelle model are in good agreement with measurements in the near hub region, which are overpredicted by the simulations without the nacelle model.  At $3D$ downstream from the turbine, the $k$ profiles computed from all the simulations are in good agreement with measurements. At $4D$, $5D$ and $6D$ downstream from the turbine, the $k$ profiles computed from the simulations with the actuator surface nacelle model are in good agreement with the measured values. The $k$ computed from the simulations without the actuator surface nacelle, on the other hand, is smaller than that from experiments at these locations. It is noticed that $3D$, $4D$ and $5D$ downstream from the turbine are the locations where maximum $k$ in the streamwise direction is observed. For further turbine downstream locations at $7D$, $8D$ and $9D$, the $k$ profiles computed from all the simulations show good agreements with measurements, although the $k$ computed from the simulations without a nacelle model is still slightly smaller than measurements.  
%

\begin{figure}
  \centerline{\includegraphics[width=0.7\textwidth, height=\textheight,keepaspectratio]{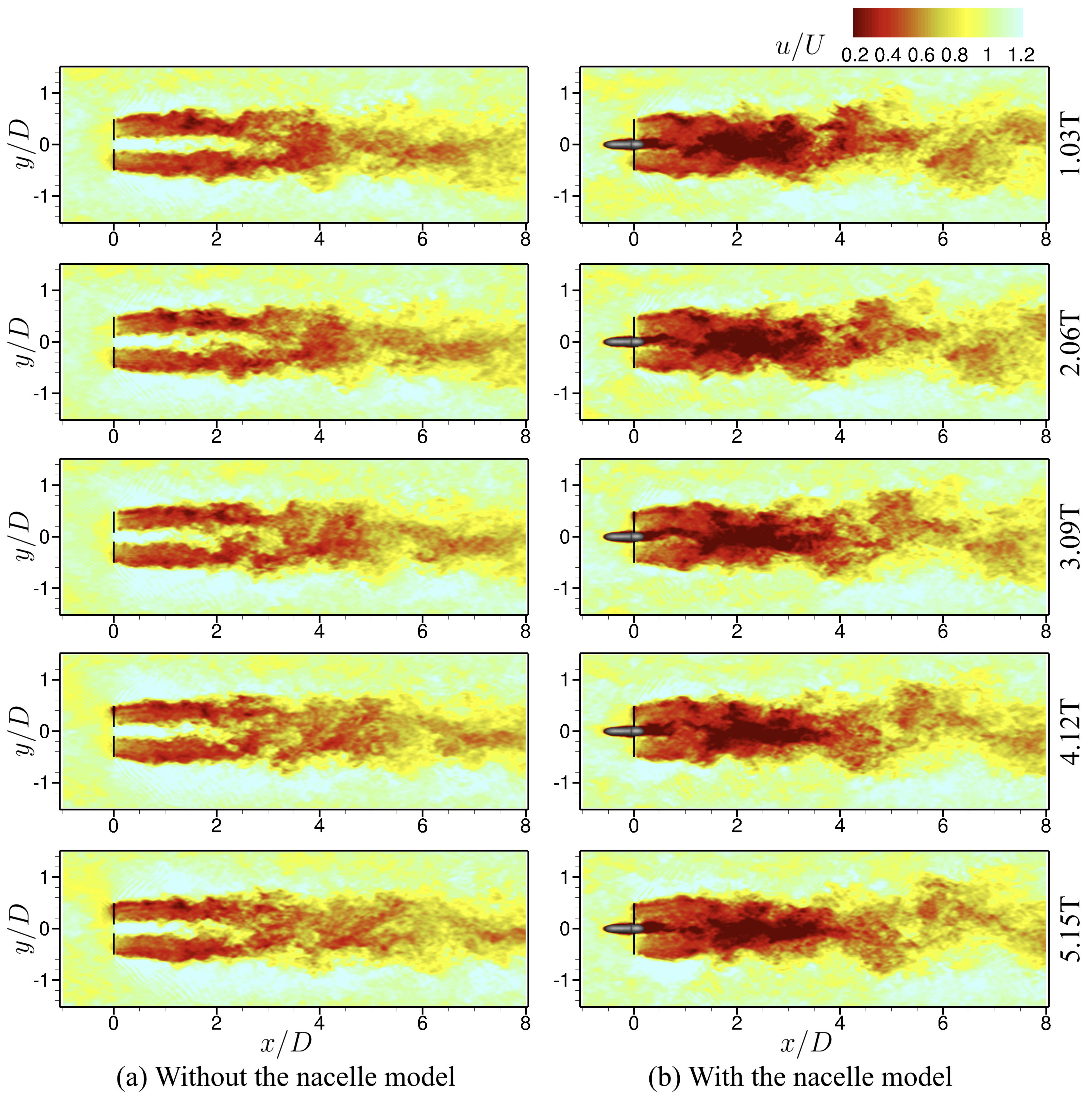}}
  \caption{Contours of instantaneous streamwise velocity at different time instants. For both simulations, the grid spacing is $D/50$. }
\label{fig:uinst_mhk}
\end{figure}

%We have shown the capability of the developed actuator surface models in capturing the turbulence kinetic energy profiles at different downstream locations. Without a nacelle model, the turbulence kinetic energy is underpredicted significantly in the region with high turbulence intensity. 
In order to show the different wake meandering behavior for the simulations with and without a nacelle model more intuitively, we show in Fig.~\ref{fig:uinst_mhk} the contours of instantaneous streamwise velocity at different time instants. It is seen that the jet from the hub of the turbine remains columnar until $3D$ turbine downstream for the ASB simulation. The inner wake from the ASB-ASN simulation, on the other hand, expands radially from $2D$ turbine downstream and interacts with the outer tip shear layer for about $1D$ to $2D$ distance.  

\begin{figure}
  \centerline{\includegraphics[width=0.8\textwidth, height=\textheight,keepaspectratio]{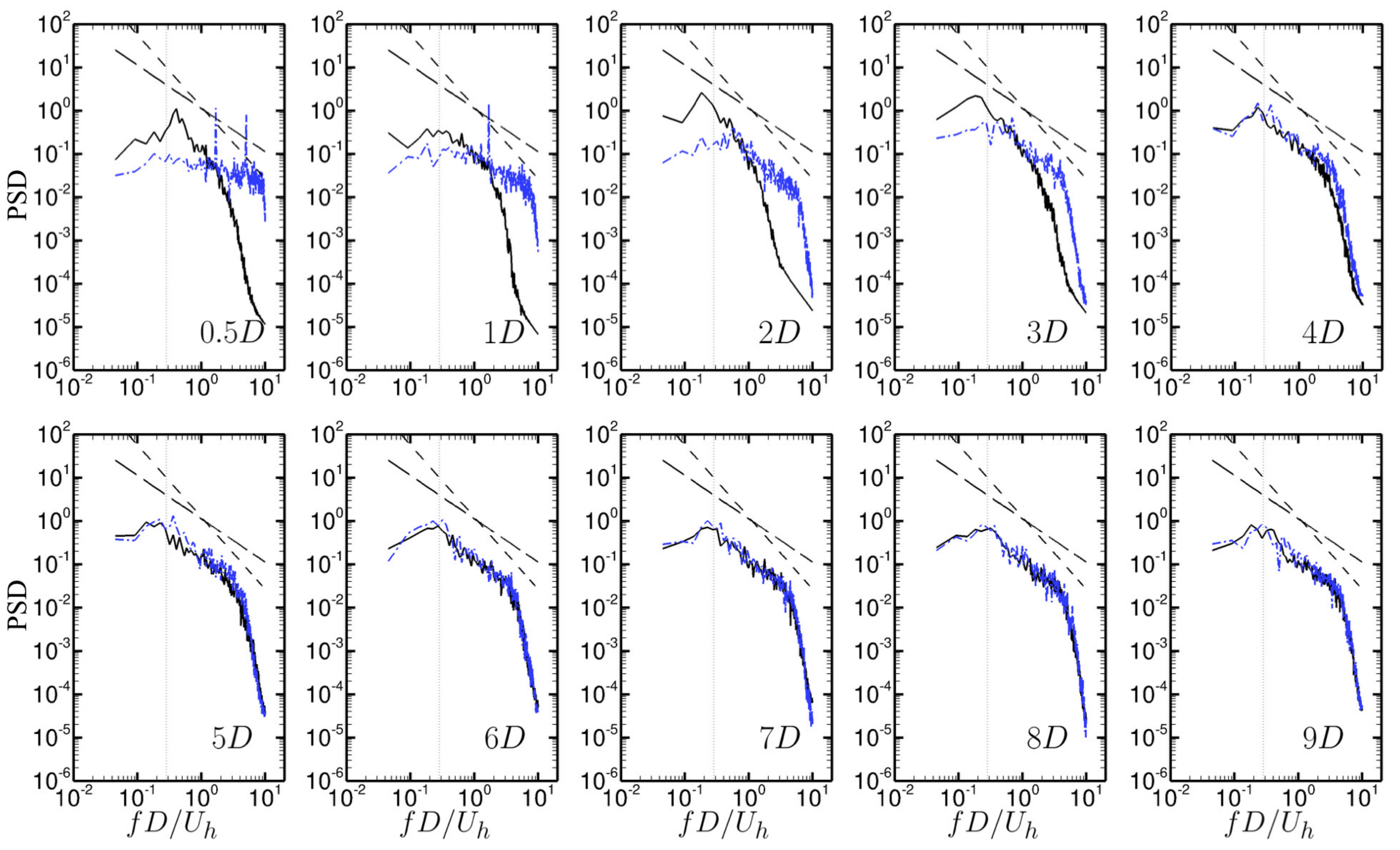}}
  \caption{Power spectral density (PSD) of spanwise velocity fluctuations along the hub of the turbine at different downstream locations for the hydrokinetic turbine case. Black solid line: ASB-ASN; Blue dash-dot line: ASB. In both cases, the grid spacing is $D/50$. PSD is normalized using the incoming streamwise velocity at turbine hub height $U_h$ and the rotor diameter $D$. For the vertical dotted line, $fD/U$ is equal to 0.28. The slope of the black dashed line is -5/3. The slope of the long dashed line is -1. }
\label{fig:PSD_MHK}
\end{figure}

To further examine the differences between the wake meandering characteristics predicted by the ASB and ASB-ASN simulations, the power spectral densities at different downstream locations are compared in Fig.~\ref{fig:PSD_MHK}. At $0.5D$ downstream from the turbine higher power spectral density is observed at low frequencies for the ASB-ASN simulation in comparison with ASB simulation (which is approximately ten times larger at the dominant frequency $fD/U=0.41$).  The rotor frequency and the blade passing frequency are observed clearly on the PSD profile computed from the ASB simulation, which are not observed for the ASB-ASN simulation for which the inner wake behind the turbine is dominated by the wake from the nacelle. At $1D$ turbine downstream, only the rotor frequency is noticeable from the PSD computed in the ASB simulation. The peak frequency at around $fD/U=0.4$ observed at $0.5D$ turbine downstream in the ASB-ASN simulation does not exist anymore at $1D$ turbine downstream. While the PSD at low frequency from the ASB-ASN simulation is still significantly larger than that from the ASB simulation. At $2D$ turbine downstream, the differences of the PSD magnitudes at low frequencies between the ASB-ASN and ASB simulations becomes even larger in comparison with that at $0.5D$ and $1D$ turbine downstream. The wake meandering frequency of this turbine is $fD/U=0.28$ as reported in Chamorro et al.'s paper~\cite{chamorro2013interaction}. At $2D$, $3D$, $4D$, $5D$ and $6D$ turbine downstream locations, similar meandering frequency around 0.2 is well predicted by the ASB-ASN simulation. Such frequency, on the other hand, is observed at much further turbine downstream locations at $4D$, $5D$ and $6D$ for the ASB simulation. At $3D$ turbine downstream location, the PSD at low frequency is still significantly larger for the ASB-ASN simulation in comparison with the ASB simulation. It is also noticed that PSD at high frequency, on the other hand, is larger for the ASB simulation. At $4D$ and further downstream locations, the differences between ASB-ASN and ASB simulations become very small for the whole spectrum. 

\begin{figure}
  \centerline{\includegraphics[width=3.5in]{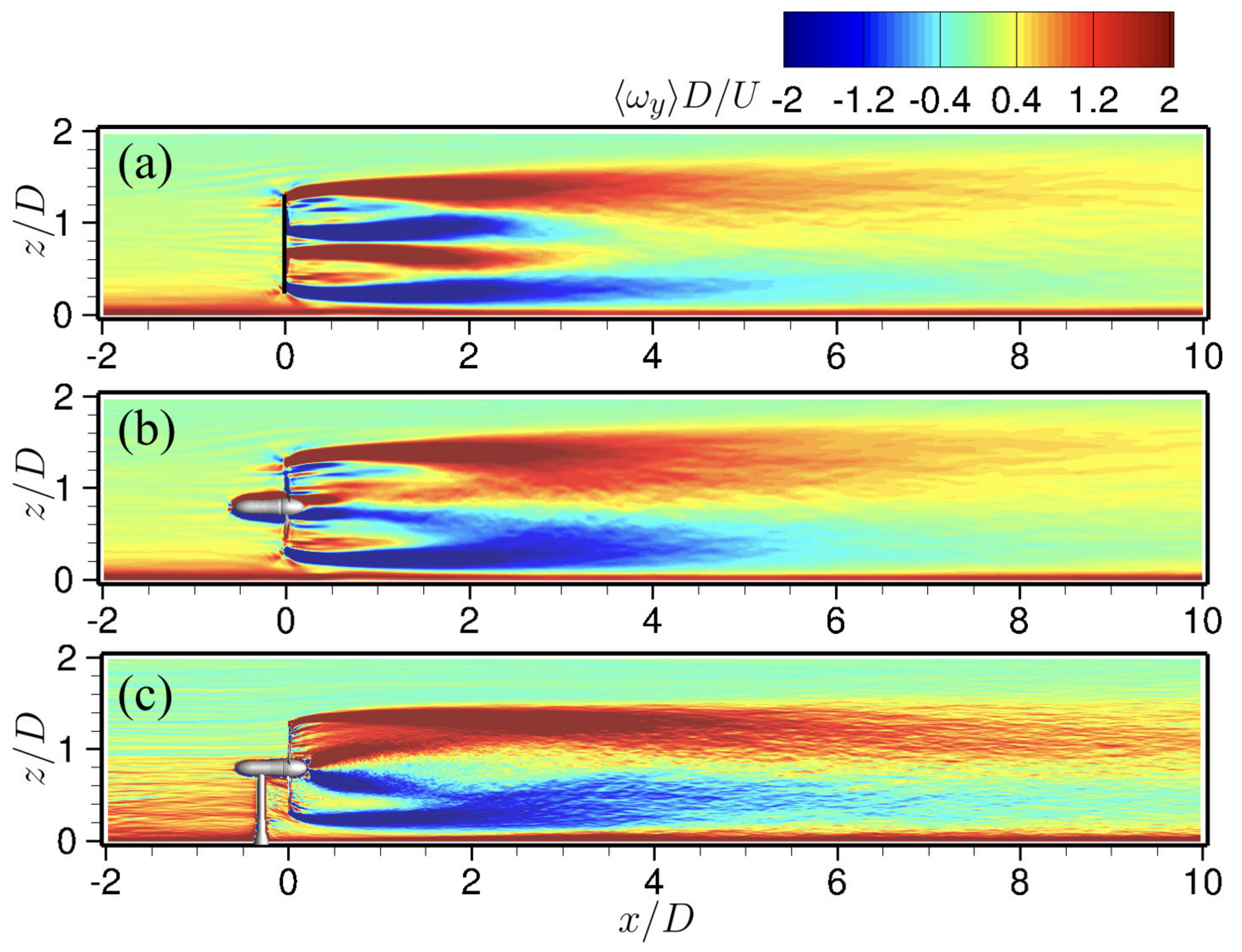}}
  \caption{Time-averaged spanwise vorticity on the vertical plane passing through the rotor center for (a) ASB and (b) ASB-ASN simulations with grid spacing $D/50$ and (c) Geometry-resolving simulation with grid spacing $D/200$~\cite{kang2014onset}. }
\label{fig:xvort_avg}
\end{figure}

Fig.~\ref{fig:uinst_mhk} has shown that the ASB-ASN model is able to predict the intensive interaction between the inner wake and outer shear layers, which plays an important role in triggering the wake meandering at far wake locations. To more clearly demonstrate the interaction between the inner wake and the outer wake, we plot in Fig.~\ref{fig:xvort_avg} the time-averaged spanwise vorticity for (a) and (b) from the actuator surface simulations, and (c) from the geometry-resolving simulation of Kang, Yang and Sotiropoulos~\cite{kang2014onset}. As seen, the sign of the vorticity within the outer tip shear layers is opposite from that within the corresponding inner shear layers for the ASB simulation as shown in Fig.~\ref{fig:uinst_mhk}(a). The vortices within the inner region remain strong until $3D$ $\sim$ $4D$ turbine downstream, which prohibits the radially expansion of the outer wake to the wake center. On the other hand, the sign of the vorticity within the outer tip shear layers is the same as that within the corresponding inner shear layers for the ASB-ASN simulation and the geometry-resolving simulation as shown in Fig.~\ref{fig:uinst_mhk}(b) and (c), respectively.  The vortices within the inner region gradually expands and joins with the vortices from the outer layers at approximately $2D$ turbine downstream, for which it happens somewhat earlier for the geometry-resolving simulation. Such interactions result in wider high-magnitude vorticity regions from approximately $2.5D$ to $6D$ turbine downstream for the simulation with the nacelle model, where the turbulent kinetic energy is underpredicted by the ASB simulation as shown in Fig.~\ref{fig:tkeprof_mhk_ASB}. Because of the different flows within the inner region, the radial expansions of the outer shear layers arrive the wake center at different downstream locations, which are approximately $3D$ (which is nearly the same for the geometry-resolving simulation as shown in (c)) and $4D$ for the simulations with and without a nacelle model. It is noticed that these two locations approximately coincide with the locations where the wake meandering frequency starts to be picked up for the ASB-ASN and ASB simulations, respectively. 
\section{Conclusions}
\label{sec:Conclusions}
In this paper, a new class of actuator surface models for wind and hydrokinetic turbines is proposed, which can take into account more geometrical features of turbine blade and nacelle as compared to the standard actuator line model. In the actuator surface model for blade, the blade is represented by the surface formed by foil chord lines at different radial locations. The forces on the blade are calculated using the blade element approach and are uniformly distributed in the chordwise direction. In the actuator surface model for nacelle, the nacelle is represented by the actual geometry of the nacelle. The normal component of the forces on the nacelle surface is calculated in the same way as in the direct forcing immersed boundary method. The magnitude of the tangential force is calculated using a friction coefficient and a reference velocity, in which the friction coefficient is computed using Schultz and Grunow's formula~\cite{schlichting2003boundary}, and the magnitude of the reference velocity is equal to the instantaneous streamwise incoming velocity upstream of the nacelle. The direction of the tangential force is taken the same as the local tangential velocity relative to the nacelle surface.  It is noticed that the proposed actuator surface model for nacelle can be applied to other bluff bodies.  Instead of using Schultz and Grunow's formula, the friction coefficient may also be set to a constant value when the Reynolds number effect is insignificant or the friction drag is negligible in comparison with the form drag. 
For complex geometries when Schultz and Grunow's relation fails, more accurate distributions of the friction coefficient  need to be employed, which can be obtained from wind tunnel experiments or wall-resolved large-eddy simulations. 

The actuator surface model for nacelle is evaluated by simulating the flow over periodic placed nacelles. The model predicted results are compared with that from the wall-resolved large-eddy simulation with the nacelle geometry represented by the sharp-interface immersed boundary method~\cite{gilmanov2005hybrid, ge2007numerical}. The comparison shows relatively large discrepancies at near wake locations, where the wake recovery rate and the turbulence intensity computed from the model are lower than the wall-resolved large-eddy simulation predictions. At far wake locations, on the other hand, overall good agreement is obtained. 

The capability of the actuator surface models in turbine wake predictions is evaluated by simulating the MEXICO turbine and the hydrokinetic turbine employed in~\cite{kang2014onset}. Three different tip-speed ratios are considered in the MEXICO turbine simulations. The comparison of the computed results with the wind tunnel measurements shows that the proposed actuator surface models can accurately predict the strength of the tip vortices and the velocity in the wake of the MEXICO turbine. The capability of the actuator surface models for wake meandering predictions is examined using the hydrokinetic turbine case. The comparison of the computed results with measurements shows that the actuator surface models can accurately predict the velocity deficits in the near wake, the turbulence kinetic energy and the characteristic wake meandering frequency in the far wake.  The importance of the actuator surface model for nacelle in the wake meandering prediction is examined by another simulation with only the actuator surface model for blade. It is shown that the simulation with only the actuator surface model for blade underpredicts the turbulence kinetic energy at approximately $4D\sim7D$ turbine downstream locations and fails to predict the wake meandering frequency at $2D$ and $3D$ turbine downstream locations. 

It is noticed that the present actuator surface model for blade is different from the actuator surface model developed by Shen, Zhang and S{\o}rensen~\cite{shen2009actuator}, in which the forces on the blade are computed from the pressure coefficients on blade surfaces and are non-uniformly distributed in the foil chordwise direction. Shen et al.'s model is more accurate in representing chordwise force distribution. To achieve this more accurate representation, it requires considerably higher spatial and temporal resolution, which may not be suitable for wind farm simulations. 
\textcolor[rgb]{0,0,0}{For the present actuator surface model, although the computational cost related to velocity interpolation (Eq.~(\ref{eq:vel-intp})) and force distribution (Eq.~(\ref{eq:force-dist})) is significantly higher than that of actuator line simulations, the computational time added by actuator surface representation (Eqs.~(\ref{eq:vel-intp}) and (\ref{eq:force-dist})) still only occupies a very small percentage of the total computational time. }
Using a background grid with resolution comparable to that in actuator line simulations, the present model provides an affordable way for field-scale wind farm simulations taking into account some geometrical effects of nacelle and blades.

%It is also noted that the present model is different from the model by Jha et al.~\cite{jha2014guidelines}, in which the chord distribution along the blade is approximated using an elliptic spanwise distribution of the Gaussian radius $\epsilon$. 

%\appendix
%\section{List of symbols}
%
\ack This work was supported by Department of Energy DOE (DE-EE0002980, DE-EE0005482 and DE-AC04-94AL85000), Sandia National Laboratories and Xcel Energy through the Renewable Development Fund (grant RD4-13). Computational resources were provided by Sandia National Laboratories, National Renewable Energy Laboratory, and the University of Minnesota Supercomputing Institute. 

\bibliographystyle{wileyj}
\bibliography{ActuatorSurface}

\begin{thebibliography}{10}
\providecommand{\url}[1]{\texttt{#1}}
\providecommand{\urlprefix}{URL }
\expandafter\ifx\csname urlstyle\endcsname\relax
  \providecommand{\doi}[1]{doi:\discretionary{}{}{}#1}\else
  \providecommand{\doi}{doi:\discretionary{}{}{}\begingroup
  \urlstyle{rm}\Url}\fi

\bibitem{schlichting2003boundary}
Schlichting H, Gersten K. \emph{Boundary-layer theory}. Springer Science \&
  Business Media, 2003.

\bibitem{tennekes1972first}
Tennekes H, Lumley JL. \emph{A first course in turbulence}. MIT press, 1972.

\bibitem{calaf2010large}
Calaf M, Meneveau C, Meyers J. Large eddy simulation study of fully developed
  wind-turbine array boundary layers. \emph{Physics of Fluids (1994-present)}
  2010; \textbf{22}(1):015\,110.

\bibitem{yang2012computational}
Yang X, Kang S, Sotiropoulos F. Computational study and modeling of turbine
  spacing effects in infinite aligned wind farms. \emph{Physics of Fluids
  (1994-present)}  2012; \textbf{24}(11):115\,107.

\bibitem{yang2014large}
Yang D, Meneveau C, Shen L. Large-eddy simulation of offshore wind farm.
  \emph{Physics of Fluids (1994-present)}  2014; \textbf{26}(2):025\,101.

\bibitem{sorensen2002numerical}
S{\'o}rensen JN, Shen WZ. Numerical modeling of wind turbine wakes.
  \emph{Journal of fluids engineering}  2002; \textbf{124}(2):393--399.

\bibitem{yang2015large}
Yang X, Sotiropoulos F, Conzemius RJ, Wachtler JN, Strong MB. Large-eddy
  simulation of turbulent flow past wind turbines/farms: the virtual wind
  simulator ({VW}i{S}). \emph{Wind Energy}  2015; \textbf{18}(12):2025--2045.

\bibitem{shen2009actuator}
Shen WZ, Zhang JH, S{\o}rensen JN. The actuator surface model: a new
  navier--stokes based model for rotor computations. \emph{Journal of Solar
  Energy Engineering}  2009; \textbf{131}(1):011\,002.

\bibitem{froude1889part}
Froude R. On the part played in propulsion by differences of fluid pressure.
  \emph{Transactions of the Institute of Naval Architects}  1889;
  \textbf{30}:390--405.

\bibitem{masson2001aerodynamic}
Masson C, Sma{\"\i}li A, Leclerc C. Aerodynamic analysis of hawts operating in
  unsteady conditions. \emph{Wind Energy}  2001; \textbf{4}(1):1--22.

\bibitem{wu2011large}
Wu YT, Port{\'e}-Agel F. Large-eddy simulation of wind-turbine wakes:
  evaluation of turbine parametrisations. \emph{Boundary-Layer Meteorology}
  2011; \textbf{138}(3):345--366.

\bibitem{chamorro2010effects}
Chamorro LP, Port{\'e}-Agel F. Effects of thermal stability and incoming
  boundary-layer flow characteristics on wind-turbine wakes: a wind-tunnel
  study. \emph{Boundary-Layer Meteorology}  2010; \textbf{136}(3):515--533.

\bibitem{hansen2015aerodynamics}
Hansen MO. \emph{Aerodynamics of wind turbines}. Routledge, 2015.

\bibitem{snel1993sectional}
Snel H, Houwink R, Bosschers J, Piers W, Van~Bussel G, Bruining A.
  \emph{Sectional prediction of 3-D effects for stalled flow on rotating blades
  and comparison with measurements}. Netherlands Energy Research Foundation
  ECN, 1993.

\bibitem{shen2005tip}
Shen WZ, Mikkelsen R, S{\o}rensen JN, Bak C. Tip loss corrections for wind
  turbine computations. \emph{Wind Energy}  2005; \textbf{8}(4):457--475.

\bibitem{shen2005tip2}
Shen WZ, S{\o}rensen JN, Mikkelsen R. Tip loss correction for
  actuator/navier--stokes computations. \emph{Journal of Solar Energy
  Engineering}  2005; \textbf{127}(2):209--213.

\bibitem{kang2014onset}
Kang S, Yang X, Sotiropoulos F. On the onset of wake meandering for an axial
  flow turbine in a turbulent open channel flow. \emph{Journal of Fluid
  Mechanics}  2014; \textbf{744}:376--403.

\bibitem{tossas2016wind}
Tossas LAM, Stevens RJ, Meneveau C. Wind turbine large-eddy simulations on very
  coarse grid resolutions using an actuator line model. \emph{34th Wind Energy
  Symposium, AIAA SciTech}  2016; :AIAA 2016--1261.

\bibitem{viola2014prediction}
Viola F, Iungo GV, Camarri S, Port{\'e}-Agel F, Gallaire F. Prediction of the
  hub vortex instability in a wind turbine wake: stability analysis with
  eddy-viscosity models calibrated on wind tunnel data. \emph{Journal of Fluid
  Mechanics}  2014; \textbf{750}:R1.

\bibitem{howard2015statistics}
Howard KB, Singh A, Sotiropoulos F, Guala M. On the statistics of wind turbine
  wake meandering: An experimental investigation. \emph{Physics of Fluids
  (1994-present)}  2015; \textbf{27}(7):075\,103.

\bibitem{foti2016wake}
Foti D, Yang X, Guala M, Sotiropoulos F. Wake meandering statistics of a model
  wind turbine: Insights gained by large eddy simulations. \emph{Physical
  Review Fluids}  2016; \textbf{1}(4):044\,407.

\bibitem{sotiropoulos2014immersed}
Sotiropoulos F, Yang X. Immersed boundary methods for simulating
  fluid--structure interaction. \emph{Progress in Aerospace Sciences}  2014;
  \textbf{65}:1--21.

\bibitem{uhlmann2005immersed}
Uhlmann M. An immersed boundary method with direct forcing for the simulation
  of particulate flows. \emph{Journal of Computational Physics}  2005;
  \textbf{209}(2):448--476.

\bibitem{yang2009smoothing}
Yang X, Zhang X, Li Z, He GW. A smoothing technique for discrete delta
  functions with application to immersed boundary method in moving boundary
  simulations. \emph{Journal of Computational Physics}  2009;
  \textbf{228}(20):7821--7836.

\bibitem{du19983}
Du Z, Selig MS. A 3-d stall-delay model for horizontal axis wind turbine
  performance prediction. \emph{AIAA Paper}  1998; \textbf{21}.

\bibitem{gilmanov2005hybrid}
Gilmanov A, Sotiropoulos F. A hybrid cartesian/immersed boundary method for
  simulating flows with 3d, geometrically complex, moving bodies. \emph{Journal
  of Computational Physics}  2005; \textbf{207}(2):457--492.

\bibitem{yang2006embedded}
Yang J, Balaras E. An embedded-boundary formulation for large-eddy simulation
  of turbulent flows interacting with moving boundaries. \emph{Journal of
  Computational Physics}  2006; \textbf{215}(1):12--40.

\bibitem{ge2007numerical}
Ge L, Sotiropoulos F. A numerical method for solving the 3d unsteady
  incompressible navier--stokes equations in curvilinear domains with complex
  immersed boundaries. \emph{Journal of computational physics}  2007;
  \textbf{225}(2):1782--1809.

\bibitem{wang2011immersed}
Wang S, Zhang X. An immersed boundary method based on discrete stream function
  formulation for two-and three-dimensional incompressible flows. \emph{Journal
  of Computational Physics}  2011; \textbf{230}(9):3479--3499.

\bibitem{cabot2000approximate}
Cabot W, Moin P. Approximate wall boundary conditions in the large-eddy
  simulation of high reynolds number flow. \emph{Flow, Turbulence and
  Combustion}  2000; \textbf{63}(1-4):269--291.

\bibitem{piomelli2002wall}
Piomelli U, Balaras E. Wall-layer models for large-eddy simulations.
  \emph{Annual review of fluid mechanics}  2002; \textbf{34}(1):349--374.

\bibitem{germano1991dynamic}
Germano M, Piomelli U, Moin P, Cabot WH. A dynamic subgrid-scale eddy viscosity
  model. \emph{Physics of Fluids A: Fluid Dynamics (1989-1993)}  1991;
  \textbf{3}(7):1760--1765.

\bibitem{schepers2007model}
Schepers J, Snel H. Model experiments in controlled conditions. \emph{ECN
  Report: ECN-E-07-042}  2007; .

\bibitem{shen2012actuator}
Shen WZ, Zhu WJ, S{\o}rensen JN. Actuator line/navier--stokes computations for
  the mexico rotor: comparison with detailed measurements. \emph{Wind Energy}
  2012; \textbf{15}(5):811--825.

\bibitem{nilsson2015validation}
Nilsson K, Shen WZ, S{\o}rensen JN, Breton SP, Ivanell S. Validation of the
  actuator line method using near wake measurements of the mexico rotor.
  \emph{Wind Energy}  2015; \textbf{18}(3):499--514.

\bibitem{kang2011high}
Kang S, Lightbody A, Hill C, Sotiropoulos F. High-resolution numerical
  simulation of turbulence in natural waterways. \emph{Advances in Water
  Resources}  2011; \textbf{34}(1):98--113.

\bibitem{chamorro2013interaction}
Chamorro L, Hill C, Morton S, Ellis C, Arndt R, Sotiropoulos F. On the
  interaction between a turbulent open channel flow and an axial-flow turbine.
  \emph{Journal of Fluid Mechanics}  2013; \textbf{716}:658--670.

\end{thebibliography}

%\begin{thebibliography}{9}

%\end{thebibliography}
\end{document}